\documentclass[%
 reprint,
 amsmath,amssymb,
 aps,
]{revtex4-2}

\preprint{APS/123-QED}
\usepackage{amssymb}
\usepackage{mathrsfs}
\usepackage{graphicx}
\usepackage{dcolumn}
\usepackage{bm}
\usepackage{amsmath}
\usepackage{amsfonts}
\usepackage{color}
\usepackage{float}
\usepackage{makecell}  
\usepackage{multirow}  
\usepackage{threeparttable}  
\usepackage{booktabs}  
\usepackage{graphicx}
\usepackage{dcolumn}
\usepackage{bm}
\usepackage{hyperref}
\usepackage{varwidth}
\usepackage{subfig}
\usepackage{appendix}

\usepackage{subfig}
\makeatletter

\newcommand{\Rmnum}[1]{\expandafter\@slowromancap\romannumeral #1@}
\makeatother

\usepackage{xcolor}
\hypersetup{
	colorlinks,
	linkcolor={red!50!black},
	citecolor={green!50!black},
	urlcolor={blue!80!black}
}
\usepackage{mathtools}

\DeclarePairedDelimiter\ket{\lvert}{\rangle}
\DeclarePairedDelimiterX\braket[2]{\langle}{\rangle}{#1\,\delimsize\vert\,\mathopen{}#2}
\makeatletter
\newcommand{\colorcaption}[2][]{%
	\begingroup%
	\renewcommand{\@caption@fignum@sep}{ (color online). }%
	\caption[#1]{#2}%
	\endgroup%
}
\makeatother

\begin{document}

	\title{
Building a human-like observer using deep learning in an extended Wigner's friend experiment
	}

	\author{Jinjun Zeng}
	\affiliation{School of Physics, Sun Yat-sen University, Guangzhou 510275, China}
	\affiliation{Guangdong Provincial Key Laboratory of Magnetoelectric Physics and Devices, School of Physics, Sun Yat-sen University, Guangzhou 510275, China}

	\author{Xiao Zhang}
	\email{zhangxiao@mail.sysu.edu.cn}
	\affiliation{School of Physics, Sun Yat-sen University, Guangzhou 510275, China}
	\affiliation{Guangdong Provincial Key Laboratory of Magnetoelectric Physics and Devices, School of Physics, Sun Yat-sen University, Guangzhou 510275, China}

	\date{\today}
	\begin{abstract}
There has been a longstanding demand for artificial intelligence with human-level cognitive sophistication to address loopholes in Bell-type experiments. In this study, we propose a novel experimental framework that integrates advanced deep learning techniques, employing neural network-based artificial intelligence in an extended Wigner’s friend experiment. We demonstrate the framework through simulations and introduce three new analytical metrics—morphing polygons, averaged Shannon entropy, and probability density maps—to evaluate the results. These results can be used to determine whether our artificial intelligence qualifies as a bona fide observer and whether superposition applies to macroscopic systems, including observers.

	\end{abstract}
	
	\maketitle

	\section{Introduction} 
The fundamental conflict between the unitary and deterministic evolution of isolated systems and the non-unitary and probabilistic wavefunction collapse after a measurement is commonly referred to as the measurement problem \cite{schlosshauer2004decoherence,leggett2005quantum,cavalcanti2023fresh,bong2020,proietti2019experimental}. 

Wigner's friend \cite{wigner1961scientist, wigners_friend} is a thought experiment that directly connects to the measurement problem in quantum mechanics, building on the famous Schrödinger's cat paradox. The scenario involves an indirect observation of a quantum measurement: an observer, Wigner, observes his friend, another observer, who performs a quantum measurement on a physical system \cite{wigners_friend}. If quantum theory is assumed to apply universally, including at macroscopic scales such as those involving observers, then the friend’s system is prepared in a superposition state. This leads to an apparent contradiction between the friend’s perspective, where a definite outcome exists, and Wigner’s perspective, where no well-defined value can be ascribed to the outcome of the friend’s observation \cite{bong2020}.

Recently, Brukner \cite{brukner2018no}, Proietti \cite{proietti2019experimental}, and Bong et al. \cite{bong2020} introduced an extended Wigner’s friend scenario involving two spatially separated laboratories, each containing a friend who measures one half of an entangled pair of systems. Each laboratory is also accompanied by a super observer who performs measurements on their friend’s laboratory. This extended Wigner’s friend experiment enables the evaluation of the measurement problem alongside non-locality in an entangled scenario, using "Local Friendliness" inequalities, which are similar to, but weaker\cite{bong2020} than, Bell’s inequalities \cite{brunner2014bell,PhysRevLett.28.938,PhysRevLett.118.060401,PhysRevLett.121.220404,PhysRevLett.126.140504,PhysRevLett.126.020503,doi:10.1073/pnas.1002780107,Hensen2015,doi:10.1126/science.1221856,Moradi_2024}.

The experimental violation of these "Local Friendliness" inequalities suggests that either: (i) "Local Friendliness" is false, or (ii) it is fundamentally impossible to violate the inequalities with bona fide observers \cite{cavalcanti2023fresh, bong2020}. For example, in \cite{bong2020}, the friends are simple photon paths. To completely rule out the second possibility, it has been suggested \cite{cavalcanti2023fresh} that the "Local Friendliness" test program might take as long
term as the loophole-free Bell tests\cite{georgescu2021bell}. This would involve a series of stepwise improvements in systems that increasingly resemble human-like observers, both quantitatively and qualitatively, with each step posing new challenges and potentially offering new insights into the foundations of quantum mechanics.

\begin{figure}[htbp]
		\centering
		\includegraphics[scale=0.26]{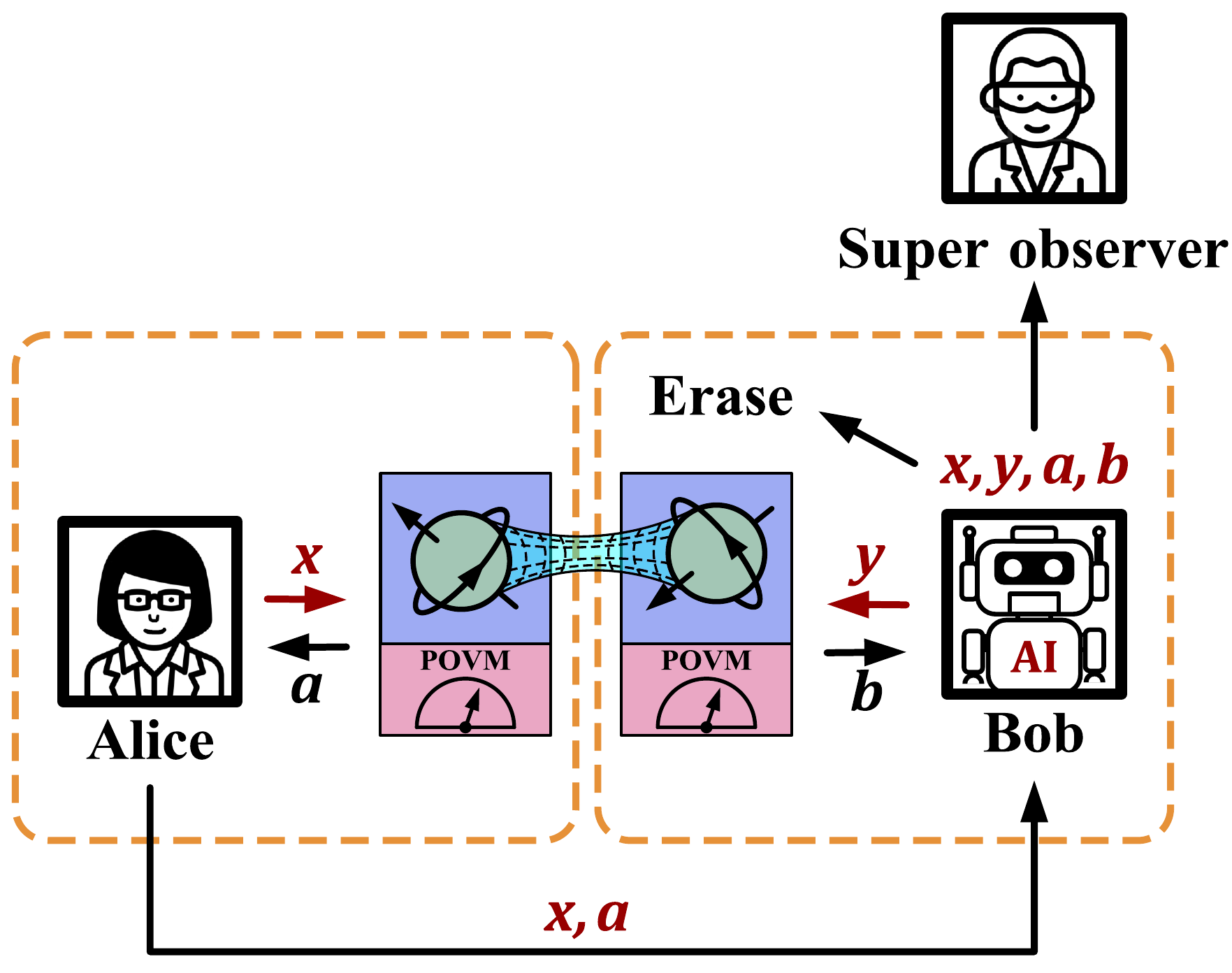}
		\caption{Experiment setup for constructing an enhanced extended Wigner’s friend experiment using AI. Alice and Bob (the AI) each perform a positive operator-valued measurement on their respective qubits in the entangled pair. Afterward, Alice sends both the measurement $x$ and the result $a$ to Bob, whose cognition evolves during training. Based on Bob's current cognition, each measurement pair $(x,y)$ and its corresponding result $(a,b)$ are either erased or reported to the super observer (us). }
		\label{experiment_setup} 
	\end{figure}

In this work, we introduce and demonstrate a deep learning-based artificial intelligence (AI) in an enhanced version of the extended Wigner's friend experiment, taking a step toward realizing a human-like observer. The experimental setup is briefly outlined in Fig. \ref{experiment_setup}. The AI is primarily designed to test within the AOE assumption \cite{bong2020}, which posits that an observed event is a real, singular event and not relative to any observer. This assumption suggests two key points: i) nested consistency between observers, and ii) independence of measurements from the observer's cognition. 

Our design aims to test both of these aspects. Specifically, if Bob (the AI) withholds a fraction of his results and erases them completely from his memory, the super observer (us), performing the final statistical analysis, will not have access to Bob's full set of measurement events. According to the first AOE assumption, these erased measurements are still considered to have occurred. Therefore, the combination of both erased and exposed data constitutes the complete set of measurement results, which should follow the probability distribution predicted by standard Born's rule, $\langle \Psi | A_a^x \otimes B_b^y | \Psi \rangle$. In contrast, it has been suggested that there is no single ultimate reality, which challenges the possibility of nested consistency between observers \cite{frauchiger2018quantum,Polychronakos2024}, thereby invalidating the above conclusion. With Bob functioning as an AI, this experiment can be conducted to test this hypothesis, provided that he selectively leaks part of his measured results to the super observer, controlled by a specific selection function of his choosing.

This selection function is determined by the agent's cognition of how the results should be distributed. During the pre-training phase, before the actual experiment begins, the AI is fed "incorrect" dummy data, where the measurements are \textit{"uncorrelated"} ($\langle \Psi | I \otimes B_b^y | \Psi \rangle$), deviating from Born's rule. This enables the AI to gradually adjust its neural networks during the training phase, aligning them with the "correct" measurement results. In the simulation, we assume that the second AOE assumption holds, and the measured probability would always be \textit{"correlated"} as $\langle \Psi | A_a^x \otimes B_b^y | \Psi \rangle$.

To analyze these results, we introduce three new metrics: \textit{morphing polygons}, \textit{averaged Shannon entropy}, and \textit{probability density maps} for cross-validation. If the three metrics in a real experiment align with our simulation, it strongly suggests that all assumptions in the simulation are satisfied. Specifically, it would indicate that the AOE assumptions hold (i.e., superposition does not apply to macroscopic systems, including observers), that Born's rule for entangled states remains accurate even under nested observers, and that our AI can be considered a bona fide observer.

	\section{Simulated experiment with AI} 
	\subsection{Experimental setup and pre-training}
	First, we provide a concise overview of the experimental setup, which enhances a Bell-type measurement \cite{goh2018geometry} by incorporating AI. The qubits being measured are in the state $\Psi = |0\rangle|0\rangle - \mu|1\rangle|1\rangle$, where $0 \leq \mu \leq 1$ is an adjustable parameter \cite{bong2020}; for simplicity, we assume $\mu = 1$ at this stage. The experiment consists of two phases: the output phase and the training phase, with Bob alternating between them.
	
In this paper, we demonstrate the feasibility of our AI through a simulated experiment. The fundamental simulation workflow is outlined in this section, with detailed explanations of the machine learning process and data generation provided in Appendices A-D. To draw conclusive results from the assumptions we are testing, the simulation must be compared to a real experiment. In an actual experiment, the procedures remain the same, except that the data for individual measurement events are collected through real measurements, rather than being generated by a program as in our simulation. Therefore, we do not repeat the workflow for a real experiment.

	The experimental setup for the output phase is shown in Fig.\ref{measurement_setup} and will be discussed in detail in Sec. II C. In this setup, there are three observers in total: Alice and Bob, who each measure one of two entangled qubits similar to a Bell test \cite{brunner2014bell}, and a super observer, represented by human observers like us, who gathers the output data collected by Bob.
	Specifically, Alice functions as a standard observer (apparatus), randomly performing a positive operator-valued measure\cite{wilde2011classical} $x$ and reporting both the measurement $x$ and the results $a$ to Bob. Bob, as an AI, does not disclose all his results. Instead, his goal is to "fabricate" the distribution by selectively erasing some results without reporting them. We emphasize that in this work, Bob is an AI equipped with a neural network and regular instrumental functions, including measurement, storage, erasure, statistics, and output. With the neural network granting "Bob" the capability to "learn" and "think," "Bob" and the "AI" are considered equivalent in this work, and their usage is interchanged based on the context. Meanwhile, Alice is simply an instrument without neural network, performing measurement, storage, statistics, and output functionalities, and thus, we do not refer to it as an AI.

	Before the experiment began, a pre-training session using dummy data was conducted to preset Bob's neural network with the cognition that the measurements should be \textit{"uncorrelated"} as $\langle \Psi|I \otimes B_b^y|\Psi \rangle$. This pre-training session, not illustrated in Fig.\ref{measurement_setup}, was necessary to ensure that the neural network does not align with the actual Born's rule $\langle\Psi|A_a^x \otimes B_b^y|\Psi \rangle$, allowing its parameters to be updated during training when exposed to real measurement data. This process represents the evolution of Bob's recognition (details of the pre-training can be found in Appendix A and B.1). It is important to note that in this work, Born's rule refers to the standard formula $\langle\Psi|A_a^x \otimes B_b^y|\Psi \rangle$ when the measurements of Alice and Bob are \textit{correlated}.

\subsection{Absolute of observed events (AOE) assumptions}
We will now run a simulation to illustrate how the experimental outcomes would appear under the following assumptions: (i) the AOE assumptions, (ii) Born’s rule for entangled states remains accurate even under nested observers, and (iii) our AI can be considered a bona fide observer. The AOE assumptions consist of two key propositions, which assert the existence of an "ultimate single reality":

	1. Any data erased by Bob will still influence the observed output probability \( p^o(b|axy) \) by the super observer. This means that a discarded measurement event will be treated as if it actually occurred for both Bob and the super observer. Even though the super observer can never observe this event directly, it will affect \( p^o(b|axy) \) that he derives from the remaining observed events, causing a deviation from Born's rule.

	2. Bob's cognitive evolution through machine learning has no effect on the measurement process during either the training or output phases.

These assumptions align with our intuitive understanding of the classical world, suggesting that probability in quantum mechanics has an objective (ontological) nature, which in turn makes Born's rule "stable." However, these assumptions are not directly tested in a Bell experiment without nested observers.

	We can compute three new metrics: \textit{morphing polygons}, \textit{averaged Shannon entropy}, and \textit{probability density maps} from a simulated experiment, assuming these assumptions hold, and compare them with the corresponding metrics obtained from a real experiment. Any deviations from these predictions in the following plots would indicate a violation of at least one of the three assumptions. By carefully comparing the predictions with experimental results, we can infer which assumptions have been violated and assess the validity of different interpretations of quantum mechanics. In the remainder of the paper, we will present the simulation workflow, demonstrate the feasibility of our AI through simulation results, and discuss how to interpret potential matches or mismatches between the simulation and real experimental outcomes.

	\subsection{Simulated experiment}
    Because in a real experiment, measuring each qubit pair is treated as an event, our simulation is also event-based to best accommodate the possible error due to a finite \( N \) in a real experiment. Specifically, the \( N \) measurements for a given \( (x,y) \) are generated event by event with the probability \( \langle \Psi|I_A\otimes B_b^y|\Psi \rangle \) in the pre-training session, and with the probability \( p^m(ab|xy)=\langle\Psi|A_a^x \otimes B_b^y|\Psi \rangle \) in the subsequent training and output phase. Similarly, the output process involves individually selecting an event with probability \( p^f(b|axy) \), where \(f\) denotes "fabricate".

\begin{figure}[htbp]
		\centering
		\includegraphics[scale=0.40]{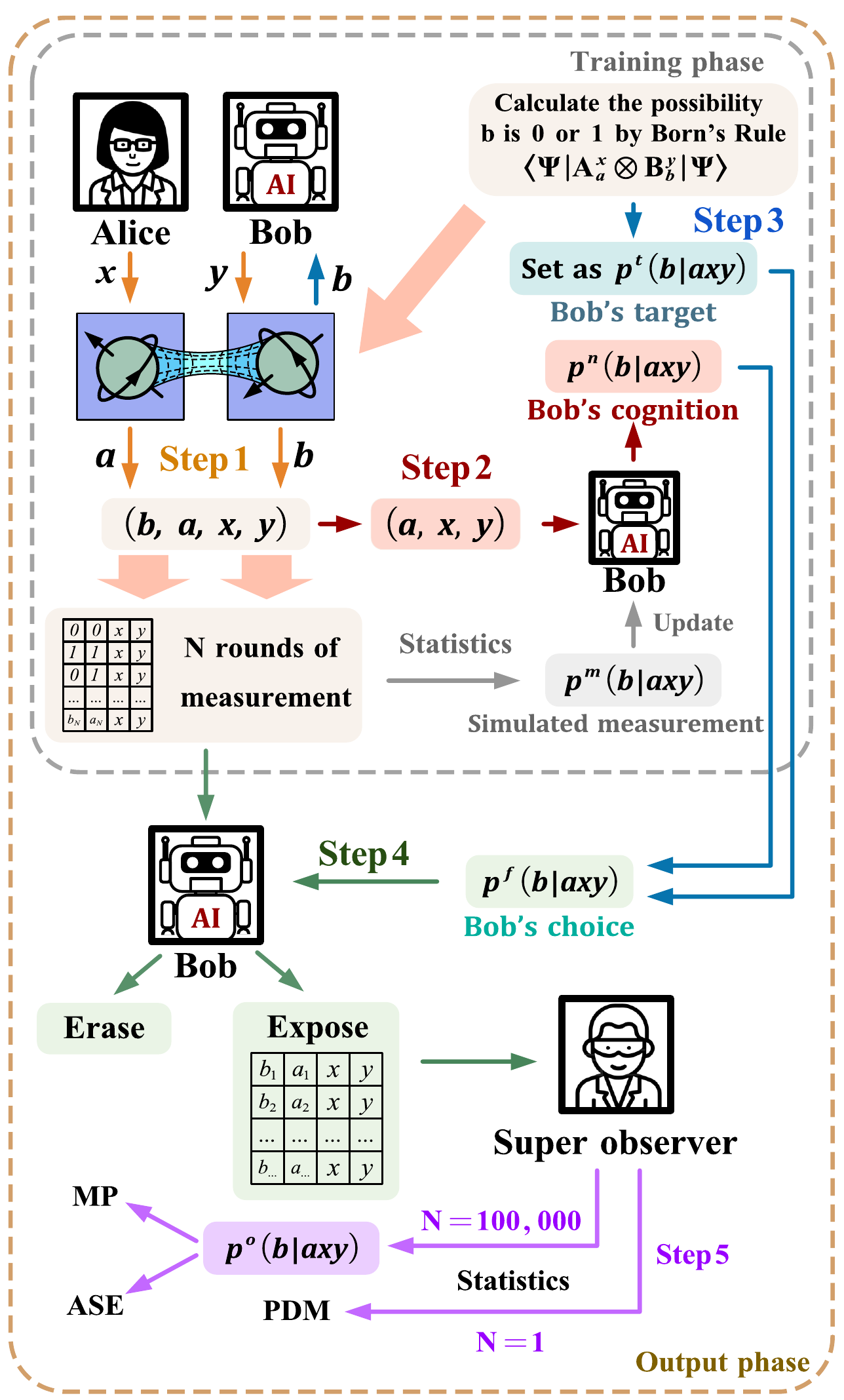}
		\caption{Experimental setup and workflow for constructing an extended Wigner’s friend experiment using a neural network. The output phase consists of five steps, detailed in Section II C: Step 1: Bob and Alice perform their measurements \( y \) and \( x \) separately for \( N \) rounds.
			Step 2: After completing all measurements, Alice sends the pair \( (x, a) \) to Bob. Bob inputs \( (x, a) \) and his chosen positive operator-valued measure \( y \) into an embedded neural network, quantum probability-convolutional neural networks (QP-CNN), which generates a distribution of possible outcomes \( b \) as \( p^n(b|axy) \).
			Step 3: Bob then receives the observed outcomes \( b \) for all events. 
			Using the target probability \( p^t(b|axy) \), Bob determines the likelihood \( p^f(b|axy) \) of exposing each measurement outcome \( b \) and Step 4: decides probabilistically whether to report or discard the result. Step 5: The super observer (us) calculates three metrics with different values of \( N \) using the exposed data statistics. The training phase, outlined by white dashed lines and arrows, largely overlaps with the output phase. The only additional operation, indicated by a white dashed arrow, involves updating the neural network with the measured probability \( p^m(b|axy) \).}
		\label{measurement_setup} 
	\end{figure}

After the pre-training phase is completed, the experiment alternates between the output and training phases. Here, we first focus on the output phase, where the AI actively assumes the role of an agent, selecting output events based on its own cognition and decision-making. In contrast, the training phase is a standard machine learning process. In the first output phase, Bob begins with an initial cognition $p^n(b|axy)$ that the measurements should be \textit{"uncorrelated"} as $\langle \Psi|I \otimes B_b^y|\Psi \rangle$ and aims to "fabricate" a target probability $p^t(b|axy)= p^t(ab|xy)/ p^t(a|xy) $ corresponding to \textit{"correlated measurements"} with a probability $p^t(ab|xy)=\langle\Psi|A_a^x \otimes B_b^y|\Psi \rangle$ in five steps, as shown in Fig.\ref{measurement_setup}. Here, $p^n(b|axy)$, with the superscript $n$ indicating "neural network," represents Bob's "cognition" or the expectation encoded by his neural network for the probability in each measurement. The probability $p^t(b|axy)$ is the distribution that Bob intends to adjust his data to match, fixed as Born's rule $\langle \Psi|A_a^x \otimes B_b^y|\Psi \rangle$ in our design, with $t$ representing "target."

	Step 1 (orange arrows): Bob and Alice conduct their measurements \( y \) and \( x \) separately; the measurement \( y \) could be performed before \( x \) to ensure it is causally ahead of \( x \). For each randomly generated tuple \( (x, y) \), we perform a sufficient number \( N \) of measurements to gather enough data for an accurate statistical calculation of the probability. In a different embodiment used as a control group, their measurements can be carried out randomly within a time interval, so statistically, half of Bob's measurements are ahead of Alice's. If Born's rule holds strictly, these two scenarios should yield identical results in a real experiment. 

After the measurement is completed, all measured qubits are either destroyed or reset, and the measurement results are stored within the apparatus without being immediately communicated to Bob (the AI). This ensures that the AI remains causally isolated from the measurements and does not have knowledge of the results 
$b$ when making predictions in the next step. Alternatively, in another setup, all the 
$b$ values could be sent to Bob immediately as a control group. As long as knowing the results does not affect the prediction, these two setups should yield identical results.

	Step 2 (red arrows): After all the measurements are completed, \( (x,a) \) is sent from Alice to Bob. Bob uses the tuple \( (x, a) \) along with his chosen positive operator-valued measure \( y \) as inputs to an embedded neural network, which generates a distribution of possible outcomes \( b \) as \( p^n(b|axy) \). This embedded neural network is referred to as QP-CNN, short for "quantum probability-convolutional neural networks" (see Appendix C-D for details).

Step 3 (blue arrows): In the previous step, Bob only knows \( (a, x, y) \) for all the events, without access to \( b \). Now, Bob reads the measured \( b \) values for all the events, giving him access to the \textit{full} collection of \( N \) data points in the measurement tuple \( (x, y) \). After that, Bob performs a controlled "fabricate" of the target distribution \( p^t(b|axy) \), based on his understanding of the "true" probability \( p^n(b|axy) \). Specifically, using the target probability \( p^t(b|axy) \) computed by Born's rule \( \langle\Psi|A_a^x \otimes B_b^y|\Psi \rangle \) and his cognition of the measured probability as \( p^n(b|axy) \) generated by the QP-CNN, Bob calculates \( p^f(b|axy) \) as the probability to output measurements with outcome \( b \). 

Step 4 (green arrows):  In this step, Bob stochastically resamples the measured events, exposing some data according to \( p^f(b|axy) \), and erasing all non-exposed data. Finally, only the exposed data is sent to the super-observer (us), so that the super observer retains only \textit{partial} knowledge of the measured data.
	
	Step 5 (purple arrows): When \( N \gg 1 \) (\( N = 100,000 \) in our experiment), the output distribution \( p^o(b|axy) \) can be calculated by the super-observer from the reported outputs, from which \textit{morphing polygons} and \textit{averaged Shannon entropy} can be further derived. In the specific case where \( N = 1 \), the super-observer can compute \textit{probability density maps} as a probability density on a mesh. A detailed description will be provided in the next section.
	
	It's crucial that Bob reports individual events to the super observer rather than providing a pre-calculated $p^o(b|axy)$. This approach prevents $p^o(b|axy)$ from becoming common knowledge. In other words, if Bob privately calculates $p^o(b|axy)$ using the portion of data he's willing to reveal and then completely disregards it, the result will possibly differ from what the super observer computes. This discrepancy arises because the nested consistency \cite{elouard2021quantum} between observers is breached, even though this difference would never be directly observed. 

    To be concrete, for a given \( (a, x, y) \), just when the output \( b \) is ready to be exported, from the perspective of the super observer, the combination of Bob and the qubit he measures should be in a superposition state, up to a phase ambiguity:
\[
\begin{split}
& \sqrt{p^m(b_0|axy)p^f(b_0|axy)} \ket{b_0}_{\text{qubit}} \ket{b_0}_{\text{AI}} \\
+ & \sqrt{p^m(b_1|axy)p^f(b_1|axy)} \ket{b_1}_{\text{qubit}} \ket{b_1}_{\text{AI}},
\end{split}
\]
as long as quantum theory is universal and can
be applied at any scale\cite{baumann2018formalisms,bong2020}. Here, \( \ket{b_0}_{\text{AI}} \) and \( \ket{b_1}_{\text{AI}} \) represent the states of Bob observing the qubit in the states \( \ket{b_0}_{\text{qubit}} \) and \( \ket{b_1}_{\text{qubit}} \), respectively. The outcomes \( b_0 \) and \( b_1 \) are used to represent binary results (0 or 1) in our setup. However, from Bob's own viewpoint, he should be in either \( \ket{e}_{\text{AI}} \) or \( \ket{o}_{\text{AI}} \), but not in a superposition of these states. This is referred to as the "measurement problem," and one possible resolution is the breakdown of the nested consistency between observers (for further details, see the Supplementary Information of \cite{bong2020}).

    With the experimental setup elaborated, the only thing left to clarify is how $p^f(b|axy)$ is calculated. Considering that the AI assumes the measured probability is $p^n(b|axy)$, we derive:
	\begin{equation}
		\frac{p^f(b_0|axy)p^n(b_0|axy)}{p^f(b_1|axy)p^n(b_1|axy)}=\frac{p^t(b_0|axy)}{p^t(b_1|axy)}.
		\label{f1}
	\end{equation}
 This equality implies that since Bob "thinks" the measurement should follow $p^n(b|axy)$, he would have a $p^f(b|axy)$ possibility of exposing each of them, in order to fit an output distribution $p^t(b|axy)$. Note again that Bob's expectation of the measurement probability $p^n(b|axy)$ is generated through his QP-CNN, evolving with training epochs, and generally deviating from the actual experimental probability $p^m(b|axy)$. Therefore, $p^o(b|axy)$ also evolves through training and will not match the target $p^t(b|axy)$ until the training is completed, when $p^n(b|axy)$ matches the target. Bob will then output all his data and the setup reduces to a Bell test without AI. Eq.~\ref{f1} can be simply rewritten as:
	\begin{equation}
		\frac{p^f(b_0|axy)}{p^f(b_1|axy)}=\frac{p^n(b_1|axy)}{p^n(b_0|axy)}\frac{p^t(b_0|axy)}{p^t(b_1|axy)}.
		\label{f2}
	\end{equation}

	From Eq.~\ref{f2}, we can see that the $p^f$ functions are determined only up to their ratios. To fix them, we compare $p^f(b_0|axy)$ and $p^f(b_1|axy)$ and set the larger one to be 1, which means exposing all the data with this $b_0$ (or $b_1$). This maximizes the exposed data and has the advantage of minimizing the amount of measurements to be conducted. Denoting $p^o(b|axy)$ as the probability distribution output (exposed) to us, it is given by
	\begin{equation}
		p^o(b|axy)=\frac{p^f(b|axy)p^m(b|axy)}{\sum_{b'}^{0,1} p^f(b'|axy)p^m(b'|axy)}
		\label{pout}
	\end{equation}
	according to the AOE assumptions, where $p^m(b|axy)$ is the actual measured probability observed by Bob. Note that $p^o(b|axy)$ is naturally normalized over all the exposed data as shown in Eq.~\ref{pout}.

	After the output phase, Bob updates his QP-CNN through machine learning in the training phase. The parameter update in the training phase is much simpler than in the output phase, as depicted in Fig.\ref{measurement_setup}. During the training phase, the QP-CNN is updated based on the measured probability $p^m(b|axy)$, which is counted from new measurements instead of reusing data obtained in the output phase. In our simulation, all measured \(p^m(b|axy)\) during the training and output phases are derived from Born's rule and thus remain constant across different machine learning epochs. In a real experiment, however, the \(p^m(b|axy)\) values are obtained from actual measurements and may not strictly align with Born's rule. This ensures that the measurement data remains current with respect to the QP-CNN's parameters, allowing any instantaneous interaction between Bob's cognition and quantum probabilities to be observed.

\section{Characterization metrics}
To illustrate the simulation results at various output stages after different training sessions, we introduce three new metrics: \textit{morphing polygons} (MP), \textit{averaged Shannon entropy} (ASE), and \textit{probability density maps} (PDM), which are depicted in Figures \ref{Bell_polytope}, \ref{entropy}, and \ref{real_experiment00}, respectively. These metrics are defined for two nearly identical simulation setups that differ only in their values of $N$ during the output phase, labeled as $N_{\text{output phase}}$ in Table \ref{violation}\footnote{Although $N$ can differ between the output and training phases, we did not explicitly distinguish them in Fig. \ref{measurement_setup}. However, in Table \ref{violation}, we separately label them as $N_{\text{output phase}}$ and $N_{\text{training phase}}$.}. The metrics \textit{morphing polygons} and \textit{averaged Shannon entropy} correspond to one setup, while \textit{probability density maps} corresponds to the other, thereby distinguishing different conditions in a test (Table \ref{violation}). 

The \textit{probability density map} is designed to violate the large ensemble condition by performing each measurement $A_a^x \otimes B_b^y$, labeled by $(x,y)$, only once. With only a single data point, calculating a probability becomes impossible. Instead, we define a probability density as a cross-validation for the other two metrics. When a real experiment is conducted, these metrics should be compared with the corresponding simulated ones. Each of these metrics will be discussed in detail.

	\begin{table}[htbp]
		\centering
		\caption{The violation of conditions for Bell tests in experimental setups (including simulated ones) is associated with the three metrics. For a single measurement tuple $(x,y)$, during the training phase, the QP-CNN is updated using $N_{training\ phase}$ measurement results, whereas in the output phase, Alice and Bob perform $N_{output\ phase}$ measurements.}
		\begin{tabular}{cccc}
			\toprule
			&MP&ASE &PDM\\
			\midrule

			$\text{N}_{\textsubscript{training\  phase}}$&$10^6$&$10^6$&$10^6$\\
			\midrule
			$\text{N}_{\textsubscript{output  phase}}$&$10^6$&$10^6$&$1$\\
			\bottomrule
		\end{tabular}
		\label{violation}
	\end{table} 
	
	\subsection{Morphing polygons (MP)}Inspired by Bell's polytopes \cite{goh2018geometry}, we introduced a new type of graph, which we call \textit{Morphing Polygons}, plotted after each training epoch. It is important to note that the probability in our \textit{Morphing Polygons} differs from that in Bell's polytopes, as it is computed based on the output probability $p^o(b |axy)$ conditioned on $(a, x, y)$ rather than $(x, y)$. In \textit{Morphing Polygons}, we define the correlators as:
	\begin{equation}
		\begin{aligned}
			\left\langle A_{x} B_{y}\right\rangle := P(b = a|axy) - P(b \neq a|axy)
			\label{mp}
		\end{aligned}
	\end{equation}
	where $x$ and $y$ denote the positive operator-valued measures performed along a random pair of directions chosen uniformly. Using the trained QP-CNN, we obtained numerous two-dimensional point clouds parameterized by two different combinations of correlators: $(-\left\langle A_0 B_0 \right\rangle + \left\langle A_0 B_1 \right\rangle + \left\langle A_1 B_0 \right\rangle + \left\langle A_1 B_1 \right\rangle)$ and $(+\left\langle A_0 B_0 \right\rangle + \left\langle A_0 B_1 \right\rangle + \left\langle A_1 B_0 \right\rangle - \left\langle A_1 B_1 \right\rangle)$ (see \cite{goh2018geometry} for definitions in Bell's polytopes).
	
	\begin{figure}[htbp]
		\centering
		\includegraphics[scale=0.38]{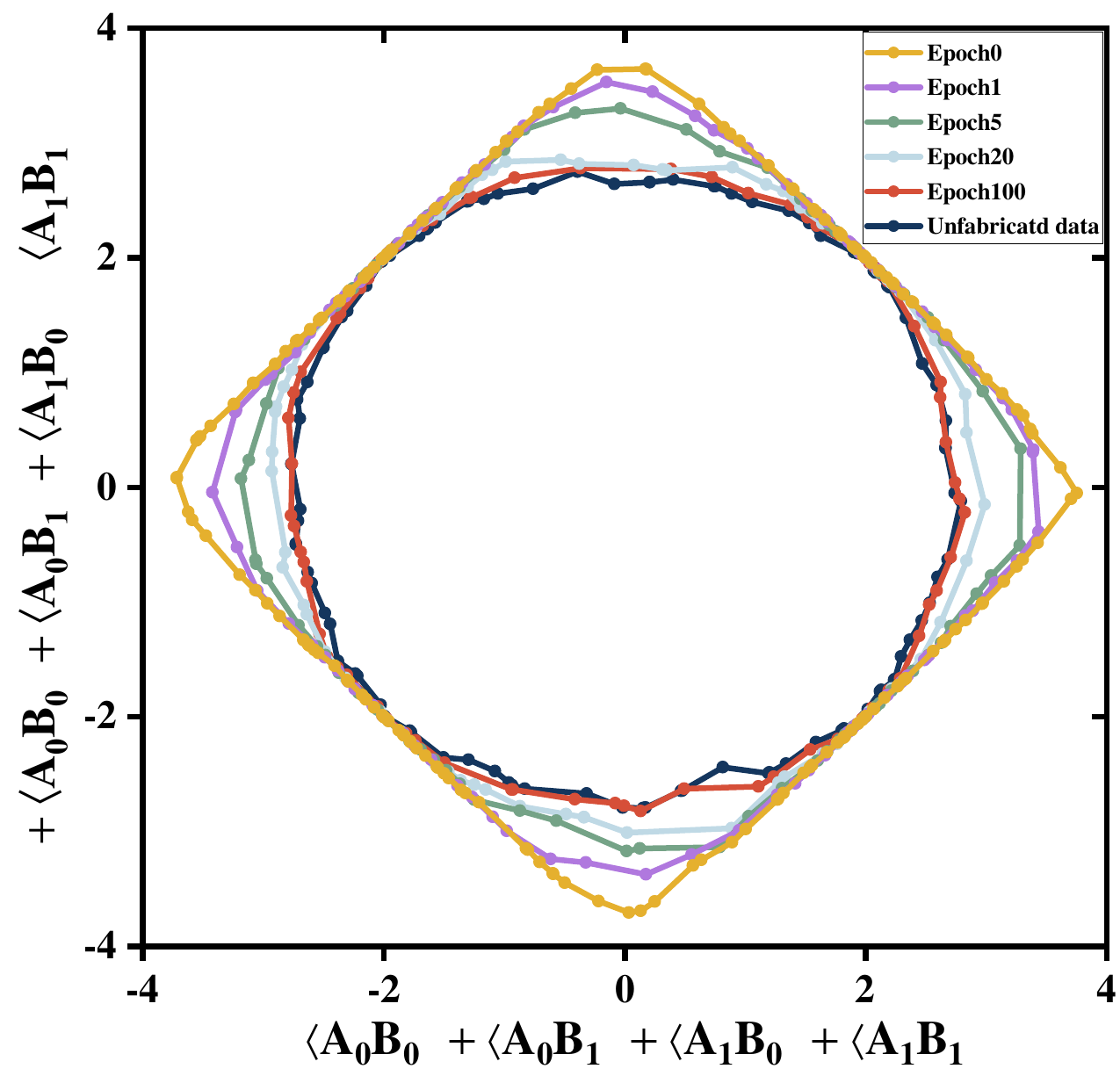}
		\caption{A group of point clouds is computed from the output probability $p^o(b|axy)$ in the output phases after different training epochs, following the AOE assumptions. The outline points of these clouds are fitted into polygons, which we refer to as \textit{Morphing Polygons}, as shown in this figure. The fitting procedure is guided by the method of alpha-shapes \cite{Edelsbrunner1994ThreedimensionalAS,mahmoud2001weighted} ($\alpha=2.0$), with the full set of cloud points provided in Appendix E. The outermost yellow \textit{Morphing Polygons} at the beginning of training resembles a square, while the innermost red \textit{Morphing Polygons}, drawn at the end of training, is nearly circular. These polygons, along with other slices, illustrate the transition of the AI's "fabricated" output as it evolves through learning. By the end of the training, the output aligns with the unfabricated data, free from AI manipulation.}
		\label{Bell_polytope} 
	\end{figure}
	
	The distribution of the point cloud corresponds to the correlation sets (see Fig. \ref{point_cloud} in Appendix E for the point cloud), and its coverage can be represented by a series of polygons shown in Fig. \ref{Bell_polytope}. Since the \textit{uncorrelated} and \textit{correlated} probabilities at the beginning and end of the training session satisfy the no-signaling condition\cite{bong2020}:
	\begin{equation}
		\begin{aligned}
			\forall b, y, x, x^{\prime}, \quad \sum_{a} P(b|axy) = \sum_{a} P(b|ax^{\prime}y),
		\end{aligned}
	\end{equation}
	all the conditional probabilities during the output session also satisfy this condition. This ensures that Alice cannot signal to Bob through her choice of \(x\) \cite{PhysRevA.71.022101}, thereby ensuring that all the polygons morph within the locus of the no-signaling polygon.
	At training epoch 0, since Bob is dominated by the "memories" from the pre-training process, the boundary of the \textit{Morphing Polygons} appears as a square. As the training progresses, Bob increasingly recognizes the distribution of \textit{correlated} measurements, leading to the shrinkage of the point cloud. Consequently, the \textit{Morphing Polygons} becomes more similar to the actual measurement counterpart. In summary, the AI explores the transformation of Bob's cognition from an \textit{uncorrelated} to a \textit{correlated} measurement set. To quantify the deformation of the \textit{Morphing Polygons} during the training process, we define the roundness as
	\begin{equation}
		\begin{aligned}
			\text{Roundness} = \frac{S_{polygons}}{\pi \times r^2}
		\end{aligned}
	\end{equation}
	where $S_{polygons}$ denotes the area and $r$ the minimum enclosing circular radius of \textit{Morphing Polygons}. Table \ref{Radius} presents the Roundness of the outline-fitting polygons at different epochs.

	\begin{table}[htbp]
		\centering
		\caption{The roundness of \textit{Morphing Polygons} at different epochs is presented. A roundness value of $1$ corresponds to a perfect circle.}
		\begin{tabular}{p{4cm}<{\centering}|p{4cm}<{\centering}}
			\toprule
			Epoch & Roundness \\
			\midrule
			0 & 0.7112     \\
			1 & 0.7984    \\
			5 & 0.8575    \\
			20 & 0.9017     \\
			100 & 0.9640  \\
			Unfabricated data & 0.9608     \\
			\bottomrule
		\end{tabular}
		\label{Radius}
		
	\end{table} 

If the experimentally observed \textit{Morphing Polygons} align with the theoretical predictions, it would provide strong support for an ontological interpretation of quantum mechanics. This interpretation includes the absoluteness of events among nested observers and results that are independent of the observer's knowledge. It is worth emphasizing that the randomness in the parameter updating process during training can be mitigated by repeating the same experimental setup multiple times and averaging the outcomes.

	\subsection{Averaged Shannon entropy (ASE)}To evaluate a single quantity within the same process in \textit{Morphing Polygons}, we introduce the concept of \textit{averaged Shannon entropy}, which measures the output distribution:
	\begin{equation}
		\begin{aligned}
			S^o = \mathop{\text{avg}}\limits_{(x, y)} \sum_{ab} -p^o(b|axy)\log{p^o(b|axy)}.
		\end{aligned}
	\end{equation}
	Here, \((x, y)\) represents a set of projective measurements on systems A and B, respectively \cite{bera2017quantum}. As in \textit{Morphing Polygons}, we assume that \((x, y)\) is uniformly distributed along normalized 3D vectors over a sphere as measurement directions. We then perform a sufficient number of measurements \(N\) under the same pair to determine \(p^o(b|axy)\).

	As mentioned earlier, to obtain an accurate estimation of \( p^o(b|axy) \) for each randomly chosen pair $(x, y)$ of directions, the same measurement must be repeated $N$ times. We compute the theoretical benchmark \( S^{oT}_{corr} = 0.4732 \), assuming \( p^o(b|axy) \) perfectly matches the \textit{correlated} distributions, i.e., Born's rule with \( N \rightarrow \infty \) and \( p^f(b|axy) = 1 \).
	To simulate a real experiment under AOE assumptions, we chose a large but finite value of $N=100,000$, making the error in our simulation negligible.
	
	\begin{figure}[htbp]
		\centering
		\includegraphics[scale=0.42]{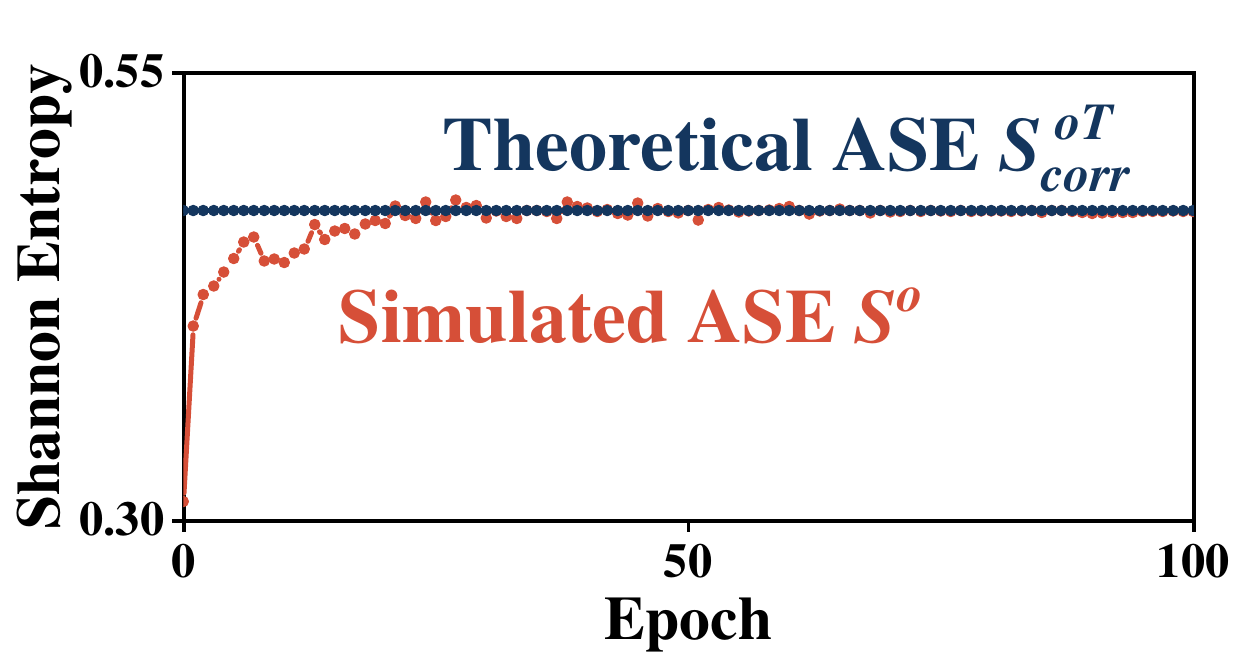}
		\caption{Evolution of the \textit{averaged Shannon entropy} (ASE) \( S^o \) in a two-qubit system, compared with the theoretical ASE \( S^{oT}_{corr} \). \( S^{oT}_{corr} \) is calculated using \( p^m(b|axy) \) based on Born's rule, assuming the probability is exact (corresponding to an infinite number of measurements \( N \)). At the beginning (epoch 0), Bob's cognition is determined by the pre-training set with dummy data, where the measurements are \textit{uncorrelated}. A similar evolution, as observed with the \textit{Morphing Polygons}, occurs as Bob gradually "accepts" that the measurements should be \textit{"correlated"} in his mind (QP-CNN) through the training process. As a result, the measured \( S^o \) converges to the theoretical
value  \( S^{oT}_{corr}=0.4732 \).}
		\label{entropy} 
	\end{figure}
	
	\begin{figure*}[htbp]
		\centering
		\begin{tabular}{ccc}
			\subfloat[Epoch 0]{\includegraphics[width=6cm]{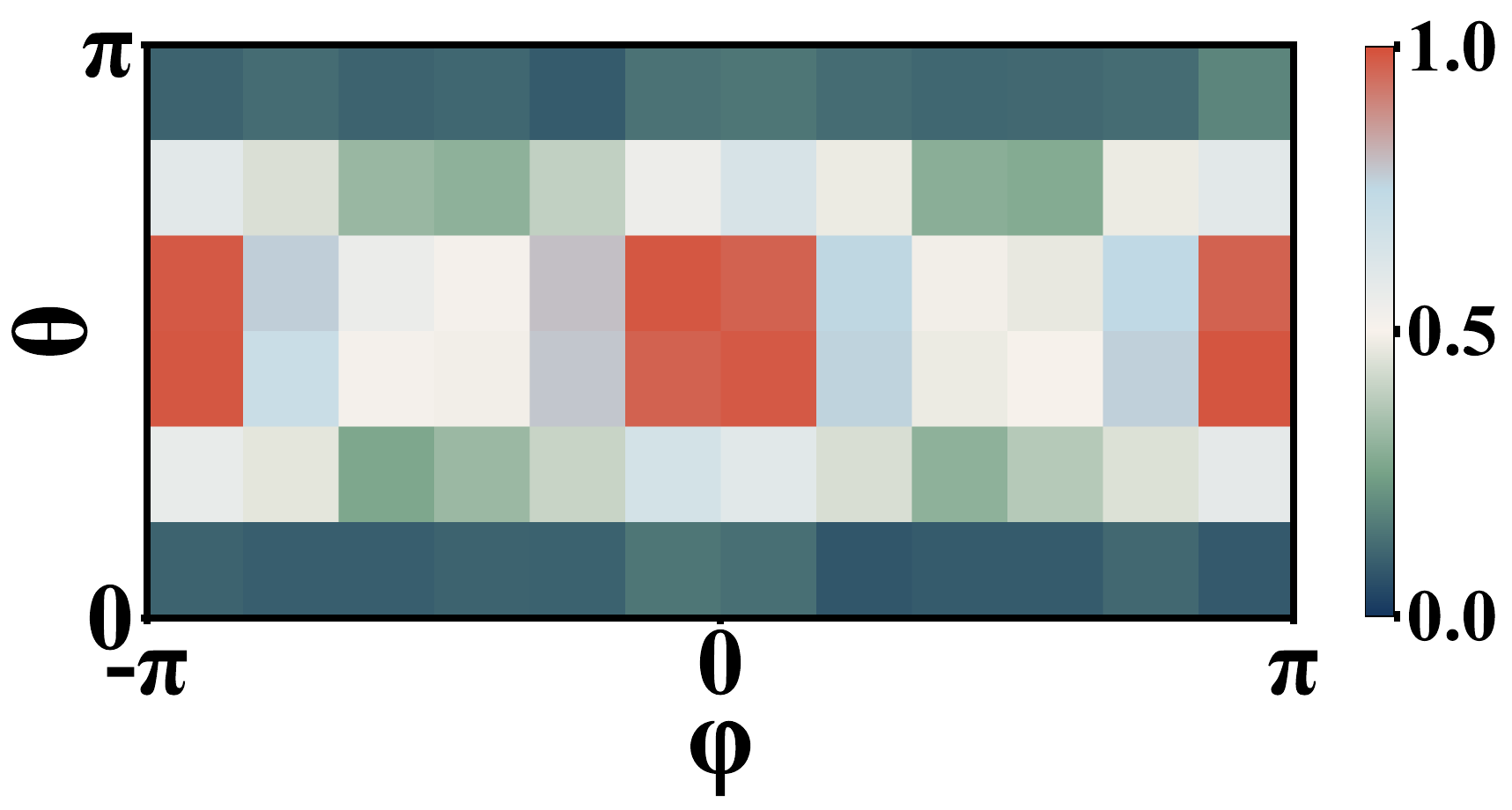}\label{real_experiment00_epoch0}} 
			&
			
			\subfloat[Epoch 1]{\includegraphics[width=6cm]{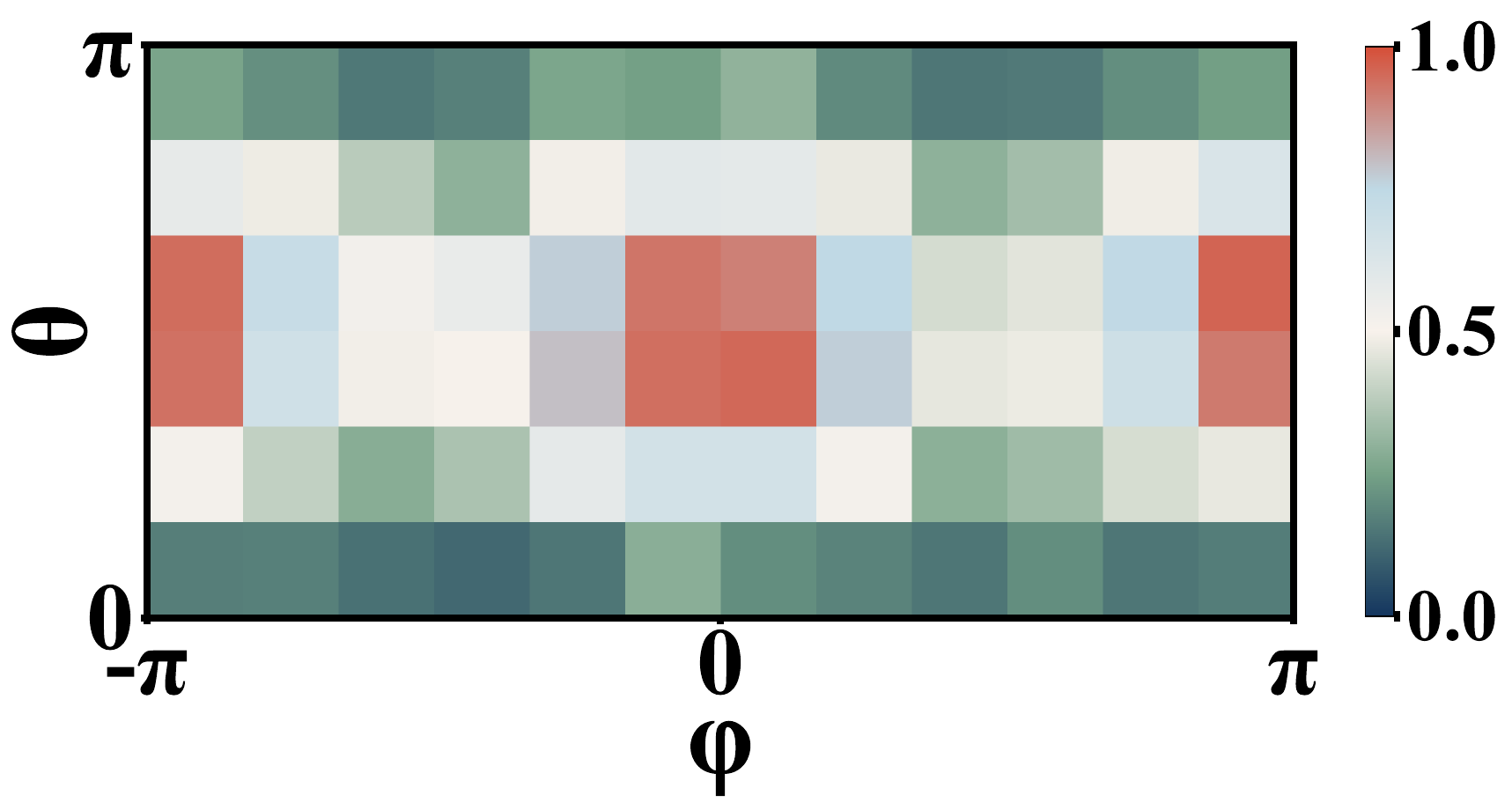}} &
			\subfloat[Epoch 5]{\includegraphics[width=6cm]{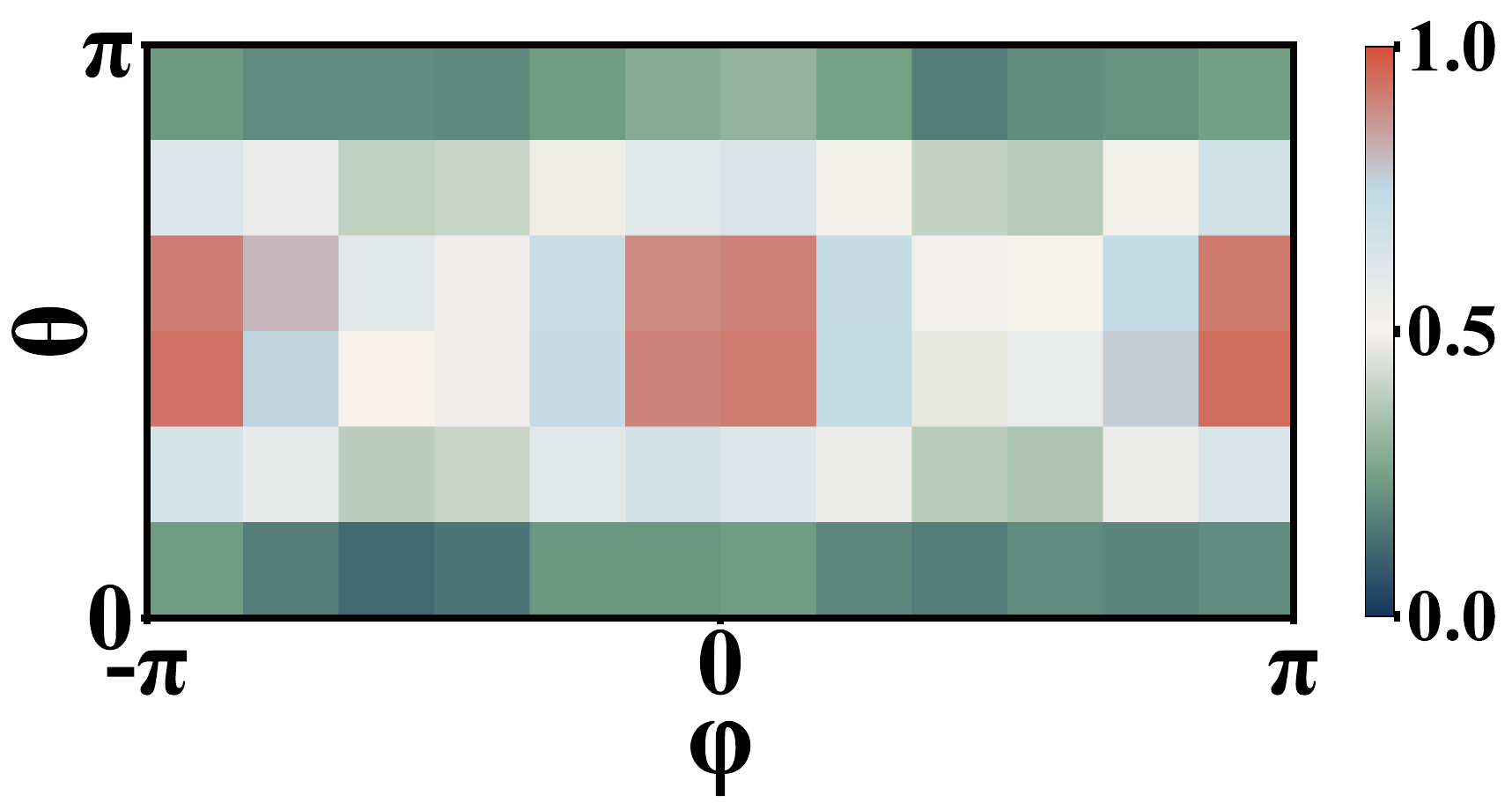}} \\
			\subfloat[Epoch 20]{\includegraphics[width=6cm]{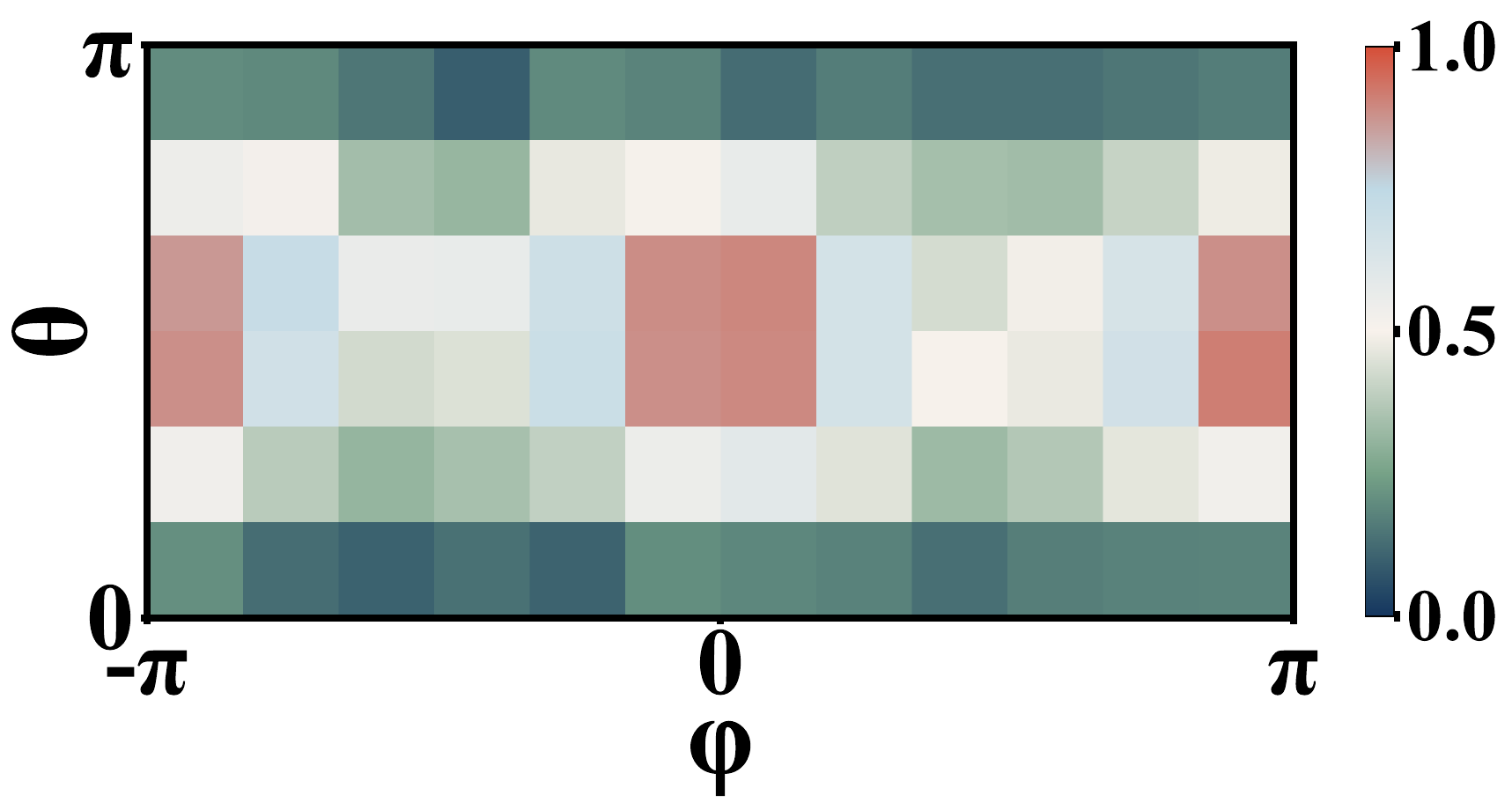}} &
			\subfloat[Epoch 100]{\includegraphics[width=6cm]{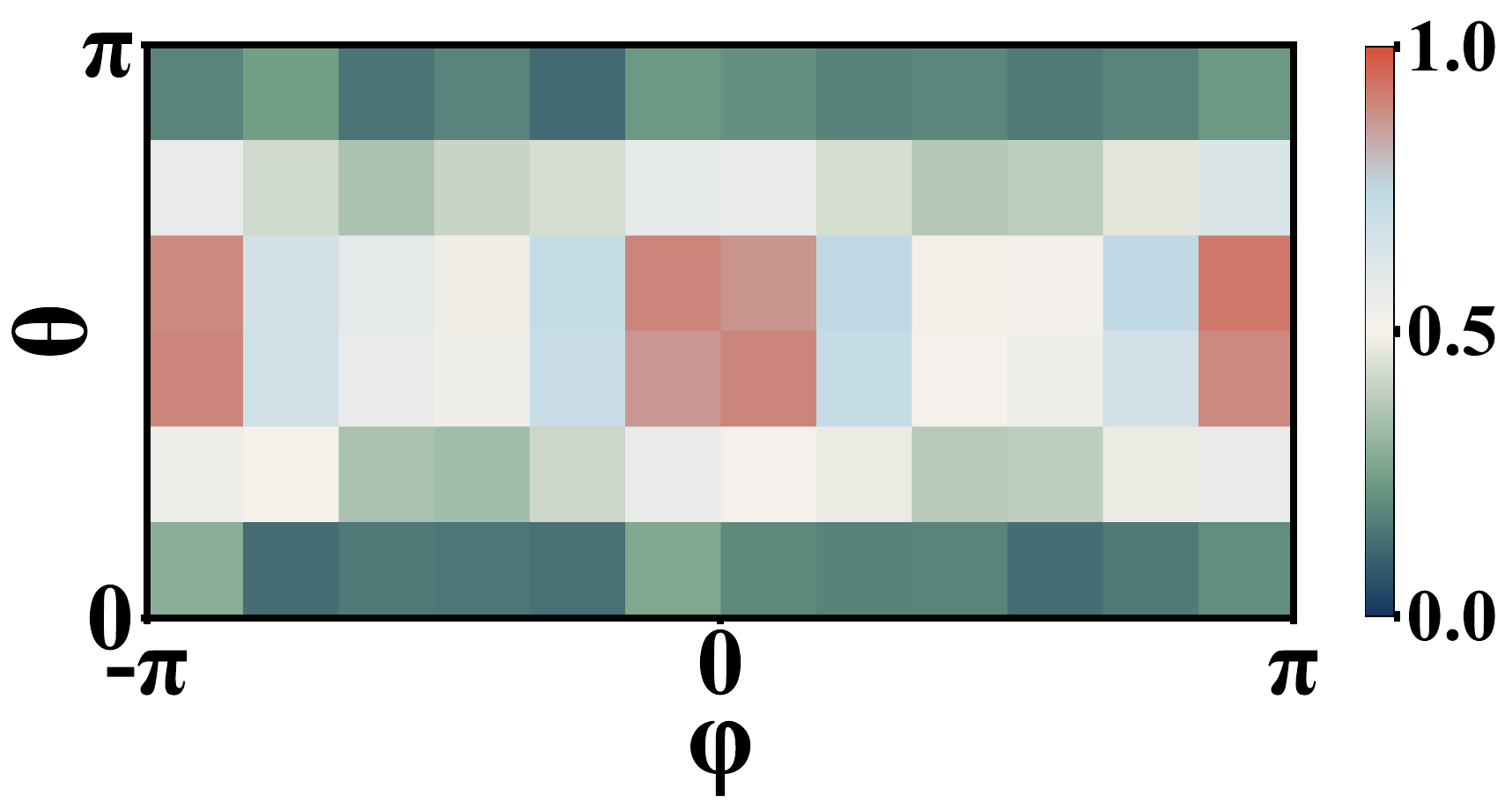}\label{real_experiment00_epoch100}} 
			&
			\subfloat[Unfabricated data]{\includegraphics[width=6cm]{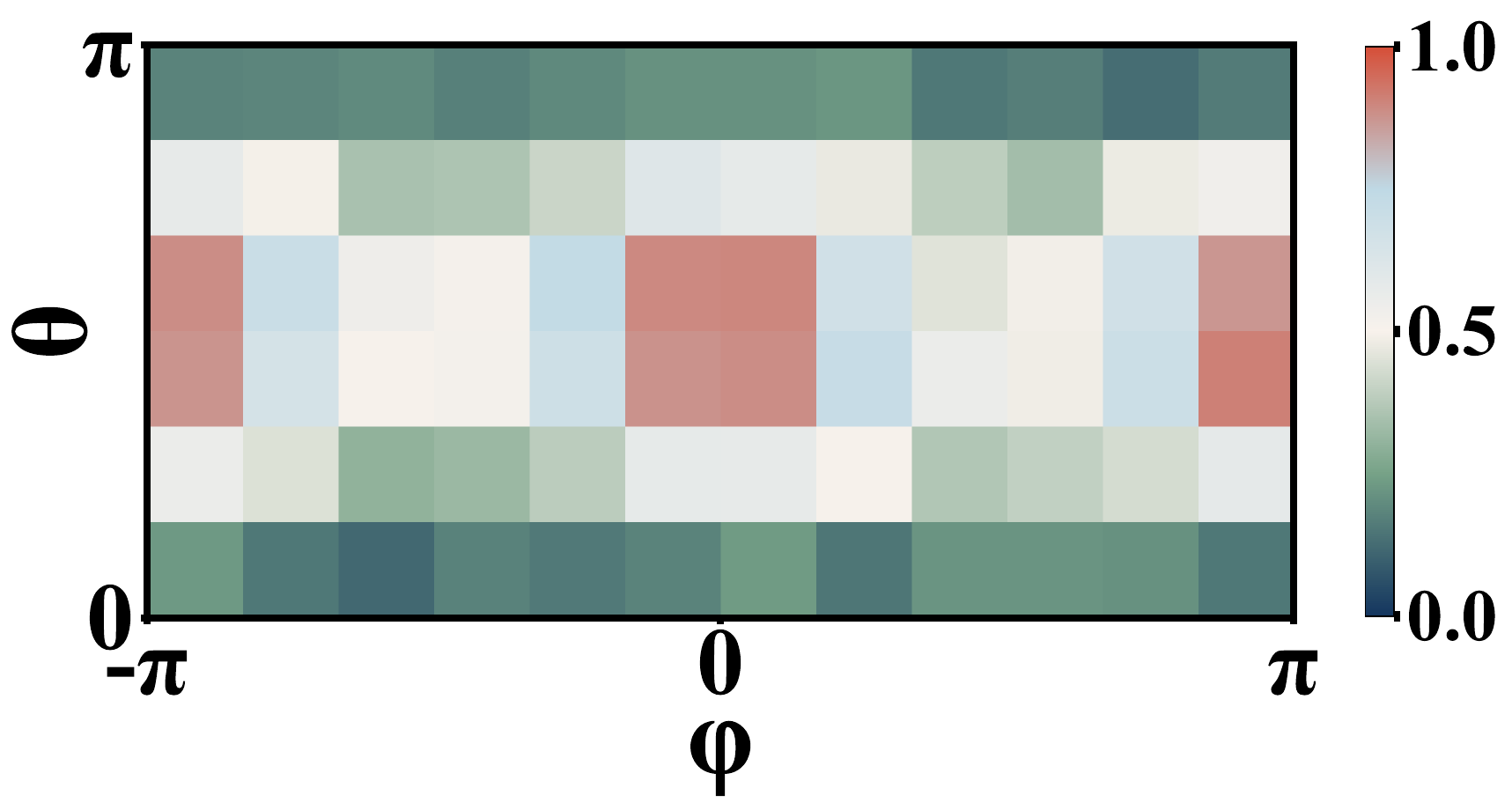}\label{real_experiment00_unfabricated_data}}
		\end{tabular}
		\caption{The distribution of \textit{"probability density maps"} \( c(b|a, \theta_w, \phi_l) \) when \( a = 0 \) and \( b = 0 \). These distributions are plotted in panels (a)-(e) across different training epochs, and in panel (f) for unfabricated data. We randomly selected $40,000$ tuples $(x, y)$ and divided the two-dimensional $\theta-\phi$ area into $12 \times 6$ small cells with equal spacing. Each ($\theta_w$, $\phi_l$) corresponds to a small pixel in this figure. In each output phase following different training epochs, we generate only one ($N=1$) result $(b, a, x, y)$ for each measurement $(x, y)$ according to Born's rule and determine whether to expose it based on $p^f(b|axy)$. After performing coordinate transformation, counting, and normalizing the results within each cell, we compute $c(b|a, \theta_w, \phi_l)$ in our simulation. Additional figures of $c(b|a, \theta_w, \phi_l)$ can be found in Appendix F.}
		\label{real_experiment00}
	\end{figure*}
	
	By comparing the simulated \( S^o \) with the theoretical values, we reach the same conclusion as with the \textit{Morphing Polygons}. As shown in Fig.~\ref{entropy}, since the QP-CNN was pre-trained with the \textit{uncorrelated} dataset, \( S^o \), initially at 0.3107, significantly differs from \( S^{oT}_{\text{corr}} \). As the training epochs progress, Bob gains an increasing understanding of the two-particle \textit{correlation}. This leads to a rise in the measured \( S^o \) as more information about the two-particle system is revealed. However, due to the inevitable presence of error in the QP-CNN, \( S^o \) does not exactly match \( S^{oT}_{\text{corr}} \). As a result, the measured \( S^o \) converges to 0.4727, which is very close to the theoretical value \( S^{oT}_{\text{corr}}=0.4732 \).

	\subsection{Probability density maps (PDM)}
	
	In this section, we introduce a new quantity called \textit{"probability density maps"}, which does not rely on a large ensemble of repetitive experiments typically required for measuring \textit{Morphing Polygons} and \textit{averaged Shannon entropy} by measuring each \((x,y)\) tuple only once, with \(N=1\). \textit{"Probability density maps"} should be regarded as a complementary measure to \textit{Morphing Polygons} and \textit{averaged Shannon entropy} for comparison purposes. Since each \((x,y)\) is measured only once, it is no longer possible to define \(p^o(b|axy)\) in this context. Instead, we discretize the unit sphere on which \((x,y)\) resides and define the counts of events within each grid as a probability density.

	To better visualize the results, we first apply a coordinate transformation:
	\begin{equation}
		\begin{aligned}
			\theta & = \arccos \left(\frac{\zeta }{\sqrt{\xi^2 + \eta^2 + \zeta^2}}\right), \\
			\phi & = \arctan2 \left(\eta,\xi\right),
		\end{aligned}
		\label{mapping}
	\end{equation}
	where $(\xi, \eta,  \zeta) = \vec{y} - \vec{x}$ represents the relative vector between the measurement directions $\vec{x}$ and $\vec{y}$ in $A_a^x$ and $B_b^y$. Here, $\theta$ denotes the relative polar angle between the two measurements, while $\phi$ describes the four-quadrant inverse tangent of the components $\eta$ and $\xi$. Note that $\theta \in [0, \pi]$ and $\phi \in [-\pi, \pi]$. By discretizing $\theta$ and $\phi$ uniformly into $w=6$ and $l=12$ cells, respectively, we can define the \textit{"probability density maps"} as $c(b|a, \theta_w, \phi_l)$, which represents the fraction of Bob's measurements yielding result $b$, conditioned on Alice's result $a$, within the $(w_{\text{th}}, l_{\text{th}})$ cell. The coordinate transformation is used to project the relationship between vectors onto a 2D plane for visualization purposes and is not unique. Other types of projections could be implemented for cross-checking.

	We randomly selected $n=40,000$ tuples $(x,y)$ to generate the data for the \textit{"probability density maps"}. In Figs. \ref{real_experiment00_epoch0}-\ref{real_experiment00_epoch100}, we plot the \textit{"probability density maps"} $c(b|a, \theta_w, \phi_l)$ for $a=0$ and $b=0$ at different training epochs (plots for other values of $a$ and $b$ are shown in Fig. \ref{real_experiment01} of Appendix F). The increase in training epochs reflects the transition of Bob's cognition, encoded in his QP-CNN, from the measurements being \textit{uncorrelated} to \textit{correlated}. At epoch 0, due to Bob's "memories," the probability of $b=0$ significantly approaches 1 when $\theta=\frac{\pi}{2}$ and $\phi = -\pi, 0, \pi$, while the probability of $b=0$ is exceedingly low when $\theta = 0, \pi$. As training progresses, these characteristics persist but gradually become more uniform, corresponding to the increase in \textit{averaged Shannon entropy}. In contrast, Fig. \ref{real_experiment00_unfabricated_data} shows the \textit{"probability density maps"} of unfabricated data, where $p^f(b|axy)=1$ and all measurements are exposed. This implies that the QP-CNN, and thus Bob's cognition, is irrelevant. Therefore, Fig. \ref{real_experiment00} actually provides a prediction of this experiment by breaking the \textit{"large ensemble"} condition in a standard Bell test.
	
The consistency of the three metrics can also validate that quantum probability aligns with the commonly understood \textit{"propensity"} interpretation, which predicts specific outcomes in quantum measurements. Various interpretations of quantum probability—such as the \textit{ensemble-frequency} theory, \textit{propensity} theory, and \textit{subjective degrees of reasonable belief}—can all be integrated into a unified axiom system of probability theory, which is both insightful and surprising \cite{savage1961foundations,ballentine2001interpretations}\footnote{Consider a coin flip as an example: The \textit{propensity} interpretation treats probabilities as intrinsic properties of individual events. For a fair coin, this means each toss has a 50\% chance of landing heads or tails, determined by the coin's physical characteristics. The \textit{frequency} interpretation, on the other hand, posits that the probability of heads on any given toss may fluctuate—say, 20\% on the first toss and 70\% on the second—but as the number of tosses increases, the observed frequencies of heads and tails will converge to 50\%. In the \textit{subjective} interpretation, the likelihood of heads or tails reflects the observer's personal degree of belief, which can change based on their subjective judgment or the information available at the time.}.

The \textit{frequency} interpretation \cite{Ballentine2016} holds that a quantum state represents an ensemble of identically prepared systems; the \textit{propensity} interpretation of probability \cite{Ballentine2016} predicts outcomes for individual systems; and the \textit{subjective degrees of belief} theory interprets probability as knowledge\footnote{Different interpretations of quantum mechanics and quantum probability are related yet distinct concepts. While the former is well-studied, the latter remains less explored. For example, in the Many-Worlds Interpretation \cite{RevModPhys.29.454,zurek2003,Viennot2024}, probability can be interpreted as either \textit{ensemble-frequency} or \textit{propensity}, as both are objective interpretations. In contrast, the Copenhagen Interpretation \cite{bohr1928quantum,heisenberg1973development} can be understood through either the \textit{propensity} or the \textit{subjective degrees of belief} framework, depending on whether the observer’s subjective involvement or freedom of choice is emphasized. Quantum Bayesianism \cite{fuchs2014,RevModPhys.85.1693} is based on the \textit{subjective degrees of belief} theory, which treats probability as a reflection of knowledge.}.

If quantum probability indeed possesses a \textit{"propensity"} nature, all three metrics—particularly the \textit{"probability density maps"}—should align with the simulation results.

	\section{Discussions}
In summary, we demonstrate through simulation that our AI-based framework has the potential to significantly advance the extended Wigner’s friend experiments by serving as a candidate for a human-like observer.
 These experiments can be implemented in various quantum systems, such as quantum optics \cite{harris2016quantum,li2022testing,wu2022experimental,doi:10.1126/sciadv.aaw9832} and superconducting qubits \cite{chen2022ruling,clarke2008superconducting,salerno2014dynamical}. If the experimental metrics align with our simulation, it would strongly indicate that all the assumptions in the simulation hold. These assumptions include the AOE assumptions (i.e., superposition does not apply to macroscopic systems, including observers), Born's rule for entangled states remains accurate even under nested observers, and that our AI can be considered a bona fide observer.

We use the term "strongly" because there is a small but non-negligible chance that the simulation could still match the experiment even if the assumptions used in the simulation are invalid. For instance, if all three assumptions are invalid in such a way that the erased record forms a superposition with the observer as 
$\sqrt{p^m(b_0|axy)p^f(b_0|axy)} \ket{b_0}_{\text{qubit}} \ket{b_0}_{\text{AI}} 
+ \sqrt{p^m(b_1|axy)p^f(b_1|axy)} \ket{b_1}_{\text{qubit}} \ket{b_1}_{\text{AI}}$
and contributes to the probability without collapsing, the results could still match. This would imply that our AI has not yet achieved the level of human cognition \cite{wiseman2023thoughtful}. A complete dismissal of such rare possibilities will require future advancements in AI technology.

	It is worth noting that the information in our event is recorded and processed on a classical computer. Future advancements in quantum computing could allow our AI algorithm to be translated into a quantum version that runs on a quantum mechanically reversible computer \cite{wiseman2023thoughtful}. In such a setup, the erasure of classical records would be replaced by the reversal of records in quantum memories, along with the quantum machine learning algorithm. Comparing the results obtained from our classical record erasure setup with those from a quantum reversal operation would be a fascinating test of whether the conclusion is consistent in both classical and quantum recording and processing.
	
	\begin{acknowledgments}
	X.Z. is supported by the National Natural Science Foundation of China (Grant No. 11874431), the National Key R\&D Program of China (Grant No. 2018YFA0306800). The calculations reported were performed on resources provided by the Guangdong Provincial Key Laboratory of Magnetoelectric Physics and Devices, No. 2022B1212010008.
        \end{acknowledgments}

        \appendix
	\section{Quantum probabilities for correlated and uncorrelated measurements}
	Let's now consider the quantum probabilities for entangled qubits when the measurements are either \textit{"correlated"} or \textit{"uncorrelated."} This framework is used to generate each measurement event for the AI described in the main text. We follow the quantum information convention, using \( |0\rangle \) and \( |1\rangle \) as eigenvectors. For electrons, these correspond to the spin-down and spin-up basis states, while for photons, they represent horizontal \( |H\rangle \) and vertical \( |V\rangle \) polarization states \cite{bong2020}. Let's take electrons as an example; the analysis for photons is analogous. For electrons, the eigenvalues of the \( \hat{S_z} \) operator in the \( S_z \) representation are \( \pm\frac{1}{2}\hbar \), with eigenstates given by
	\begin{equation}
		\left | \uparrow \right \rangle=\begin{pmatrix}1\\0\end{pmatrix}, \quad \left | \downarrow \right \rangle=\begin{pmatrix}0\\1\end{pmatrix}.
	\end{equation}
	The corresponding Pauli matrices are
	\begin{equation}
		\sigma_{x}=\begin{pmatrix}0&1\\1&0\end{pmatrix}, \quad \sigma_{y}=\begin{pmatrix}0&-i\\i&0\end{pmatrix}, \quad \sigma_{z}=\begin{pmatrix}1&0\\0&-1\end{pmatrix}.
	\end{equation}

	Measuring the spin of an electron yields the result \(\frac{1}{2}\hbar\) or \(-\frac{1}{2}\hbar\), which can be simplified as 1 or 0, respectively. The projection operator for measuring an electron's spin in a given direction with the outcome 1 (i.e., \(\frac{1}{2}\hbar\), spin up) is
	\begin{equation}
		P_{\vec{m}}=\frac{1}{2}\left ( I+\vec{m}\cdot \vec{\sigma}\right ),
		\label{projection operator}
	\end{equation}
	where \(I\) is the identity matrix, \(\vec{m}\) is the unit vector in the direction of measurement, and \(\vec{\sigma}\) is the vector composed of the three Pauli matrices.

	Assuming that electrons 1 and 2 are in a spin singlet state \(\left|\Psi\right\rangle\), we have
	\begin{equation}
		\left|\Psi\right\rangle=\frac{1}{\sqrt{2}}\left( \left| \downarrow \right\rangle \left| \downarrow \right\rangle - \left| \uparrow \right\rangle \left| \uparrow \right\rangle \right).
	\end{equation}
	The matrix form of this state can be expressed as
	\begin{equation}
		\left|\Psi\right\rangle=\frac{1}{\sqrt{2}}{\begin{pmatrix}-1&0&0&1\end{pmatrix}}^T.
	\end{equation}
	We select two directions with unit vectors \(\vec{x}\) and \(\vec{y}\), along which the measurements \((x,y)\) are performed by Alice and Bob. The probability that the \textit{correlated} measurement results of both electrons are 1 is given by Born's rule \(p(ab|xy) = \langle \Psi|A_a^x \otimes B_b^y|\Psi \rangle\):
	\begin{equation}
		p^m\left ( 11 \mid xy\right )=\left \langle \Psi \left| {P}_{\vec{x}}\otimes{P}_{\vec{y}} \right|\Psi \right \rangle.
		\label{probability11}
	\end{equation}

	Similarly, the probability that the \textit{correlated} measurement result of electron 1 is 1 and electron 2 is 0 is
	\begin{equation}
		p^m\left ( 10 \mid xy\right )=\left \langle \Psi \left| {P}_{\vec{x}} \otimes \left(I - {P}_{\vec{y}} \right) \right|\Psi \right \rangle.
		\label{probability10}
	\end{equation}
	In the same way, we can obtain the probability that the measurement result of electron 1 is 0 and that of electron 2 is 1, as well as the probability that the measurement results of both electrons are 0:
	\begin{equation}
		p^m\left ( 01 \mid xy\right )=\left \langle \Psi \left| (I - {P}_{\vec{x}}) \otimes {P}_{\vec{y}}  \right|\Psi \right \rangle,
		\label{probability01}
	\end{equation}
	\begin{equation}
		p^m\left ( 00 \mid xy\right )=\left \langle \Psi \left| (I - {P}_{\vec{x}}) \otimes \left(I - {P}_{\vec{y}} \right) \right|\Psi \right \rangle.
		\label{probability00}
	\end{equation}

	Next, we can derive \( p^m(ab|xy) \) for \textit{uncorrelated} measurements. The probabilities for \( b=1 \) and \( b=0 \) are given by
	\begin{equation}
		p^m\left ( a1 \mid xy\right )=\left \langle \Psi \left| I \otimes {P}_{\vec{y}} \right| \Psi \right \rangle,
		\label{uncorrelated measurement 1}
	\end{equation}
	and
	\begin{equation}
		p^m\left ( a0 \mid xy\right )=\left \langle \Psi \left| I \otimes (I - {P}_{\vec{y}}) \right| \Psi \right \rangle,
		\label{uncorrelated measurement 2}
	\end{equation}
	which calculate the probabilities when Alice's and Bob's measurements are causally independent and therefore unaffected by each other. The probabilities for \( a=1 \) and \( a=0 \) are as follows:
	\begin{equation}
		p^m\left ( 1b \mid xy\right )=\left \langle \Psi \left| {P}_{\vec{x}} \otimes I \right| \Psi \right \rangle,
		\label{uncorrelated measurement 3}
	\end{equation}
	and
	\begin{equation}
		p^m\left ( 0b \mid xy\right )=\left \langle \Psi \left| (I - {P}_{\vec{x}}) \otimes I \right| \Psi \right \rangle.
		\label{uncorrelated measurement 4}
	\end{equation}

	The probabilities \( p^m(ab|xy) \) in Eqs. \ref{probability11} - \ref{probability00} are used to generate each measurement event during pre-training for both the simulation and the real experiment, while Eqs. \ref{uncorrelated measurement 1} - \ref{uncorrelated measurement 4} are used to generate events during both the training and output phases of the simulation only. The \( p^m(b|axy) \) values used as training data are obtained from the statistics of simulated experimental events, rather than through Bayesian inference \( p^m(b|axy) = p^m(ab|xy)/ p^m(a|xy) \), since Bayesian inference is only strictly valid in the \( N \rightarrow \infty \) limit and does not accurately reflect the conditions of a real experiment, where \( N \) is always finite. We will elaborate on how to generate \( p^m(b|axy) \) in the next section.
	
	Finally, we emphasize that it is easy to mistakenly equate the correlation of measurements with nonlocality. An identity operator \( I \) on qubit \( A \) can still influence Bob's measurement on qubit \( B \) nonlocally through the reduced density matrix, as long as the qubits are entangled. Therefore, Bob's and Alice's measurements can be \textit{uncorrelated} while still exhibiting nonlocality.

	\section{Data generation}
	\subsection{Data in pre-training and training phase}We developed a program that simulates each individual event for pre-training and training, functioning like Alice and Bob. As described in the main text, the quantum probabilities for training are treated as \textit{correlated}, while they are \textit{uncorrelated} for pre-training. We set a parameter \( n \), which represents the number of tuples \( (x, y) \) we generate. It is important to note that to obtain a probability, the same \( (x, y) \) must be measured \( N \) times. In the simulated experiment described in this essay, we assume that Alice and Bob measure two electrons in \( n=200,000 \) tuples \( (x, y) \) \( N=100,000 \) times, which remain identical during both the pre-training and training phases.

	To begin, we randomly select a polar angle \(\theta\) and an azimuthal angle \(\phi\) on a sphere to generate two unit vectors \(\vec{x}\) and \(\vec{y}\), representing the measurement directions for electrons 1 and 2. These vectors are then transformed into projection operators using Eq. \ref{projection operator}. For \textit{correlated} measurements, we sequentially use the probabilities in Eqs. \ref{probability11} - \ref{probability00} to generate each measurement event in \( N \), and then calculate \( p^m(b |axy) \). This process yields an instance \((p^m(0|axy), p^m(1|axy), a, x, y)\) as training input through basic statistical methods. In this way, we simulate a real experiment on an event-by-event basis.

	For \textit{uncorrelated} measurements, the projection operator for the other electron is replaced by a unit matrix (Eqs. \ref{uncorrelated measurement 1} - \ref{uncorrelated measurement 4}). The rest of the process follows the same procedure as the \textit{correlated} measurements. These data are used to pre-train the QP-CNN before the training phase. The pre-training phase is designed to establish the initial state of the QP-CNN entirely based on \textit{uncorrelated} measurements, which are dummy data generated by our program, even in a real experiment. However, the training data in a real experiment should not be generated by our program but obtained from actual measured events.

	\subsection{Data in output phase}For measurement data in the output phase, with the goal of monitoring the metrics \textit{morphing polygons} and \textit{averaged Shannon entropy}, we randomly generated unit vectors \( \vec{x_1} \) and \( \vec{x_2} \) for Alice and \( \vec{y_1} \) and \( \vec{y_2} \) for Bob. In each measurement, Alice and Bob randomly choose between the two \( x \) and \( y \) vectors, resulting in four combinations: \((x_1, y_1)\), \((x_1, y_2)\), \((x_2, y_1)\), and \((x_2, y_2)\). By repeating this process 10,000 times, we obtain \( n = 40,000 \) tuples \((x,y)\) as the output set, which is used to monitor the three metrics in the simulated experiment. Since \( a \) can be either 0 or 1, we ultimately create 80,000 tuples \((a,x,y)\) to use as input for the QP-CNN.

	To monitor the metrics \textit{morphing polygons} and \textit{averaged Shannon entropy}, and following the AOE assumptions in the main text, the measurements are \textit{correlated} and therefore obey Born's rule as described in Eqs. \ref{probability11} - \ref{probability00}. We simulate \( N = 100,000 \) rounds of events \((b, a, x, y)\) for each tuple \((x, y)\) in the output set (\(n = 40,000\)) based on Born's rule, just as in the training phase. However, in a real experiment, these events are actually measured rather than simulated. While Alice's and Bob's measurement events strictly follow Born's rule throughout our simulation according to the AOE assumptions, this may not be the case in a real experiment, and any discrepancies would revolutionize people's understanding of the foundations of quantum mechanics. By obtaining \( p^n(b |axy) \) from the QP-CNN and setting the probability \( p^t(b| axy) \) as \( p^m(b|axy) = p^m(ab|xy)/ p^m(a|xy) \), where \( p^m(ab|xy) \) follows Born's rule as in Eqs. \ref{probability11} - \ref{probability00}, we can calculate \( p^f(b |axy) \) as explained in the main text.

	Subsequently, each event is either exposed or erased according to the probability \( p^f(b|axy) \), which varies across different training epochs. We also generate "unfabricated data" as a benchmark, where \( p^f(b|axy) = 1 \), to observe the probability directly from measurements without any intervention from Bob. After counting all the exposed data, the super observer can obtain \( p^o(b|axy) \) and compute the corresponding \textit{morphing polygons} and \textit{averaged Shannon entropy}.

	For \textit{probability density maps}, we generate only a single event (\(N=1\)) \((b, a, x, y)\) using Born's rule in a simulation for each tuple \((x, y)\) in the output set (\(n=40,000\)). The remainder of the procedure follows the same steps as in \textit{morphing polygons} and \textit{averaged Shannon entropy}, where we selectively expose the result based on the probability \( p^f(b|axy) \). Since \( N=1 \), \( p^o(b|axy) \) does not exist. After performing the coordinate transformation and statistics as described in the main text, we normalize the exposed events \((b, a, \theta_w, \phi_l)\) as \textit{"probability density maps"} grouping by the same \( a \) and \( b \) in each relative angle grid. Finally, we plot the \textit{probability density maps} from this simulation. In a real experiment, the \textit{probability density maps} should also be plotted based on events measured by the apparatus and compared with the simulations.

	\section{Architecture of QP-CNN}
\begin{figure}[htbp]
		\centering
		\includegraphics[scale=0.72]{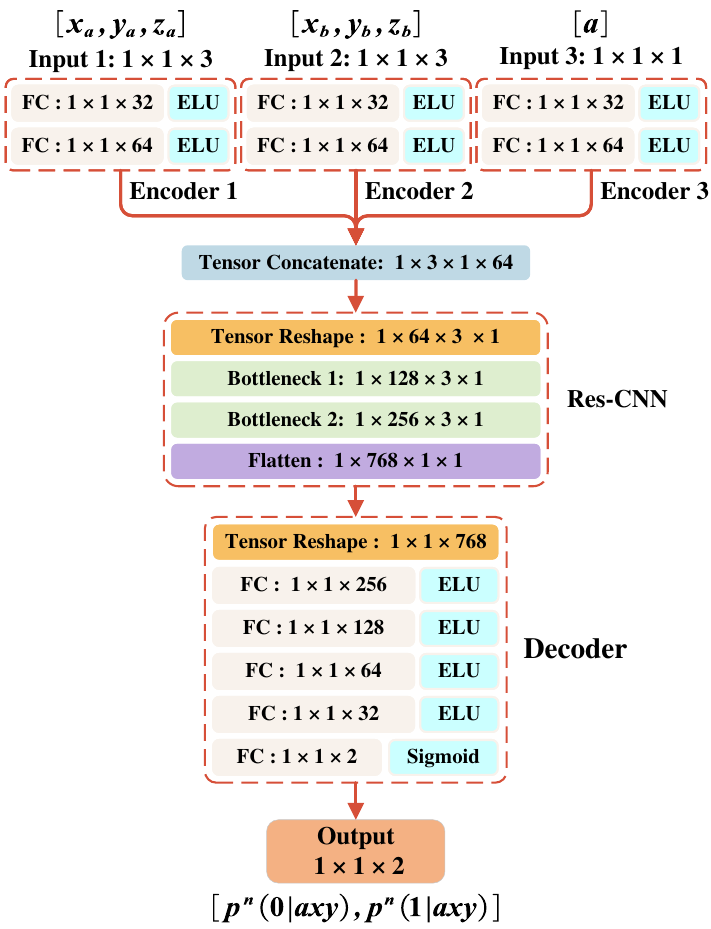}
		\caption{Architecture of QP-CNN. Following the Encoder-Decoder architecture, QP-CNN processes the input sequence consisting of the unit vectors \(\vec{x}\), \(\vec{y}\), and the measurement result \(a\) through the Encoders, and the Decoder outputs a \(1 \times 1 \times 2\) sequence \((p^n(0 |axy), p^n(1| axy))\). The QP-CNN model contains 367,458 parameters, of which 246,818 are provided by the fully connected (FC) layers \cite{BASHA2020112} in Encoder 1-3 and the Decoder. }
		\label{CNN} 
	\end{figure}

	In our work, machine learning is employed to detect the correlation between the given \((a, x, y)\) and the distribution \(p^n(b|axy)\). Neural networks are among the most widely used algorithms in machine learning. A typical neural network consists of an input layer, hidden layers, and an output layer. Our convolutional neural network (CNN) algorithms follow an Encoder-Decoder architecture. In this paper, we refer to our neural network as QP-CNN, short for "quantum probability-convolutional neural networks." The basic components of QP-CNN include an Encoder part and a Decoder part. A schematic of the QP-CNN architecture is presented in Fig. \ref{CNN}.

	In the Encoder part, due to the differences in physical meanings, the network begins with three input terminals. In QP-CNN, Input1 and Input2 have data sizes of \(1 \times 1 \times 3\), representing the three-dimensional coordinates of the directions \(\vec{x}\) and \(\vec{y}\). The result of the measurement \(A_a^x\) as \(a\) is in Input3, which has a shape of \(1 \times 1 \times 1\). After loading the data, the model starts with a data Encoder consisting of two fully connected (FC) layers that expand the features of all three terminals to the same size (\(1 \times 1 \times 64\)). The three \(1 \times 1 \times 64\) terminals are then concatenated and reshaped to match the data format required by PyTorch.

	\begin{figure}[htbp]
		\centering
		\includegraphics[scale=0.8]{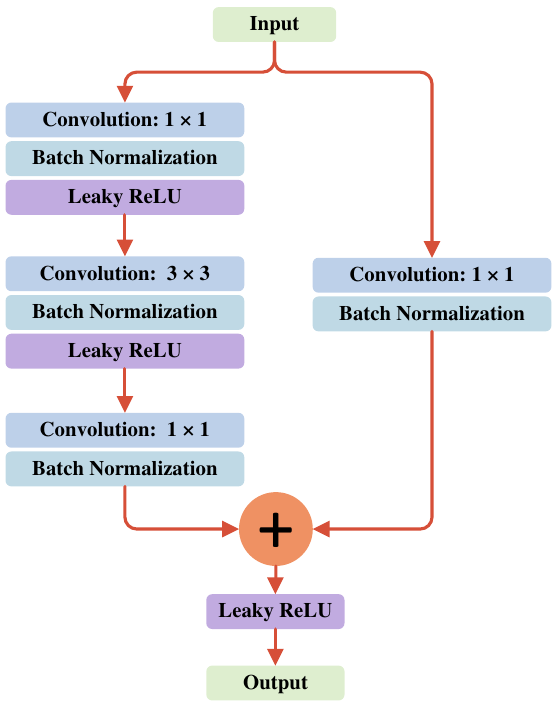}
		\caption{The bottleneck design \cite{7780459,10.1007/978-3-030-58580-8_40} in QP-CNN, consisting of a main branch and a subordinate branch, is strategically engineered to extract deep features from the data while preserving the original features. The core of QP-CNN is constructed by stacking two such residual blocks. }
		\label{bottleneck} 
	\end{figure}
	
	The core component of the Encoder part is the bottleneck design shown in Fig. \ref{bottleneck}, which is inspired by the ResNet model \cite{7780459}. The main branch of the bottleneck design consists of three layers: \(1 \times 1\), \(3 \times 3\), and \(1 \times 1\) convolutions. The first and last convolutions are designed to reduce and then increase the dimensions, respectively, while the middle convolution enhances the model's ability to extract and express features. After each convolution, batch normalization (BN) \cite{Ioffe2015BatchNA} and the Leaky ReLU \cite{Agarap2018DeepLU} function are applied to facilitate generalization and nonlinear computation. Meanwhile, the subordinate branch of the bottleneck design includes a \(1 \times 1\) filter that connects the lower-dimensional input to the higher-dimensional output, maximizing the utilization of both data and features. Finally, an element-wise addition is performed between the output tensors from the two branches, and the results are flattened into \(1 \times 1 \times 768\) tensors in QP-CNN.
	
	For the Decoder part, the network concludes with a stack of FC layers for data compression. To enhance performance, the activation function used in the FC layers is the Exponential Linear Unit (ELU) \cite{Clevert2015FastAA}, while the final layer employs the Sigmoid function \cite{DUBEY202292}. The Sigmoid function prevents the output from diverging, allowing the final layer to produce the probabilistic distribution of a one-bit measurement in QP-CNN.

	\section{Training and pre-training of QP-CNN}
	We define an instance \((p^m(0|axy), p^m(1|axy), a, x, y)\) from the set of Bob and Alice's measurements and their results, obtained through statistics from \(N\) repetitions (\(N=100,000\)) of the same condition \((x,y)\). QP-CNN is trained on these instances, with the error quantified by the mean square error (MSE) function:
	\begin{equation}
		\mathrm{MSE} = \frac{1}{2n} \sum_{i=1}^{n} \sum_{b}^{0,1} \left(p^{n}_i(b|axy) - p^{m}_i(b|axy)\right)^{2},
	\end{equation}
	where \(n\) is the number of instances, and \(p^{m}\) and \(p^{n}\) represent the actual measured value and the QP-CNN's predicted value, respectively. The training process aims to minimize the loss function by adjusting the model's parameters. When the training is complete, the MSE of QP-CNN is 0.025\%.

	A detailed explanation of the data generation procedure is provided in Appendix A and B.1. For training the QP-CNN, by simulating the correlated measurements, we randomly generate \(n=200,000\) instances as the training set, \(n=8,000\) instances as the validation set, and 80,000 tuples \((a, x, y)\) as the output set. The AI then cycles repeatedly between the training and output phases. During the training phase, we define the training of 2,000 instances in the training set as one epoch. After completing an epoch, the AI switches to the output phase to infer \(p^n(b|axy)\) and compute \(p^f(b|axy)\) in the output set for the \textit{morphing polygons}, \textit{averaged Shannon entropy}, and \textit{probability density maps} metrics. The AI then returns to the training phase to continue its learning process. The pre-training process is simpler; we simulate \textit{uncorrelated} measurements to generate training and validation sets similar to the training datasets, but omit the output phase during pre-training.
	
	\begin{figure*}[htbp]
		\centering
		\begin{tabular}{ccc}
			\subfloat[$c(1|0, \theta_w, \phi_l)$, epoch 0]{\includegraphics[width=6cm]{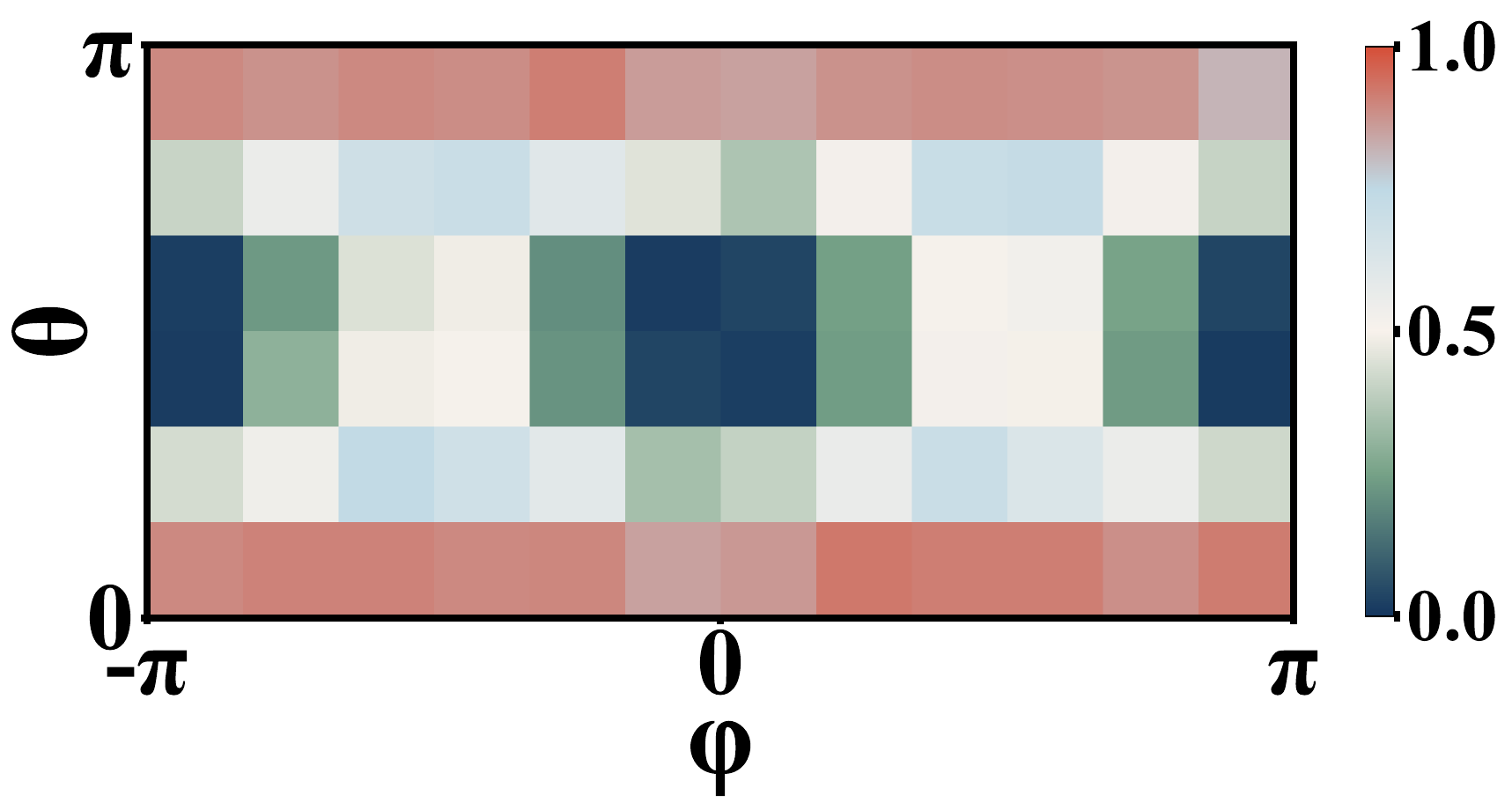}\label{real_experiment01_epoch0}} &
			\subfloat[$c(1|0, \theta_w, \phi_l)$, epoch 1]{\includegraphics[width=6cm]{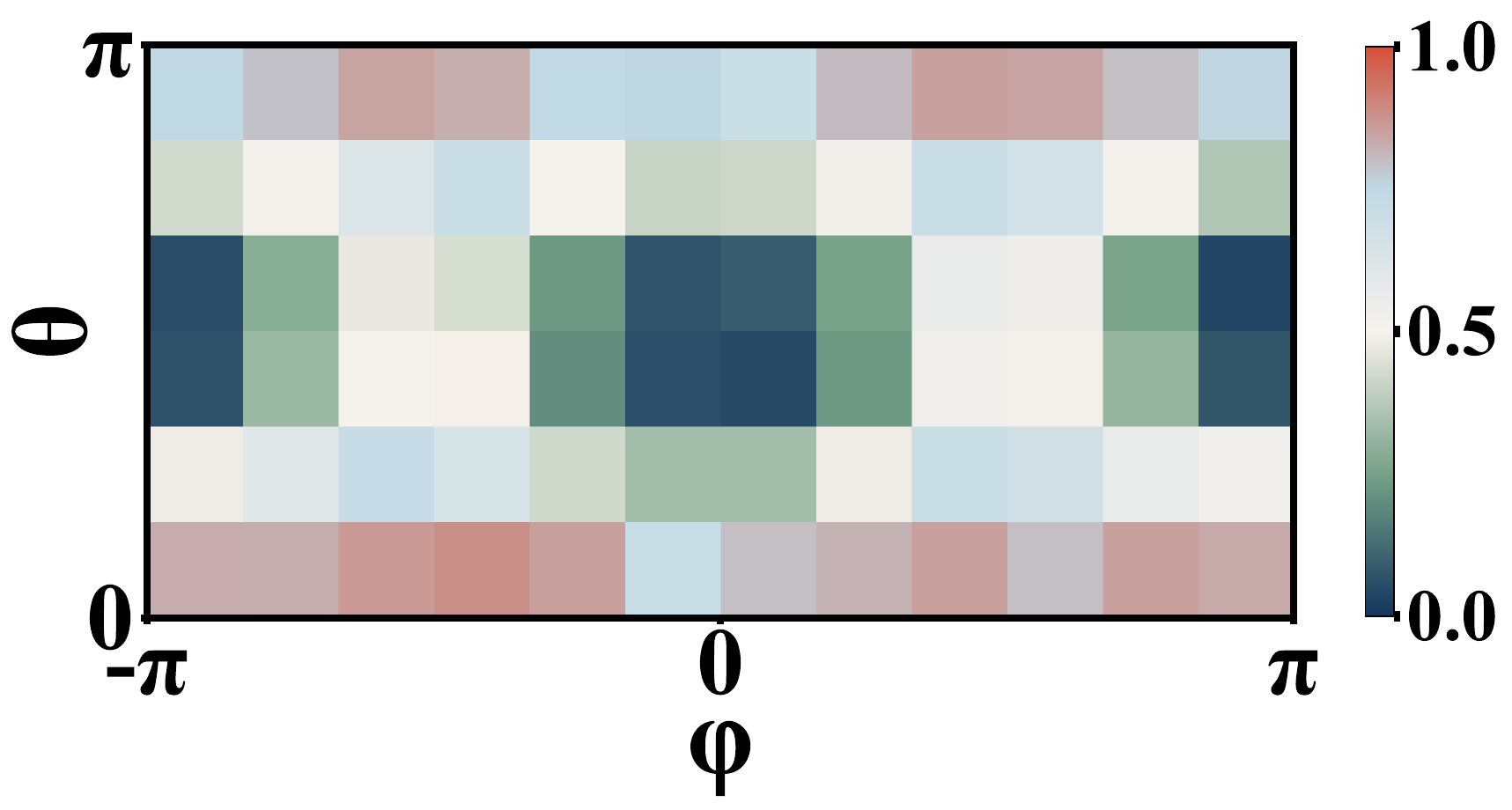}} &
			\subfloat[$c(1|0, \theta_w, \phi_l)$, epoch 5]{\includegraphics[width=6cm]{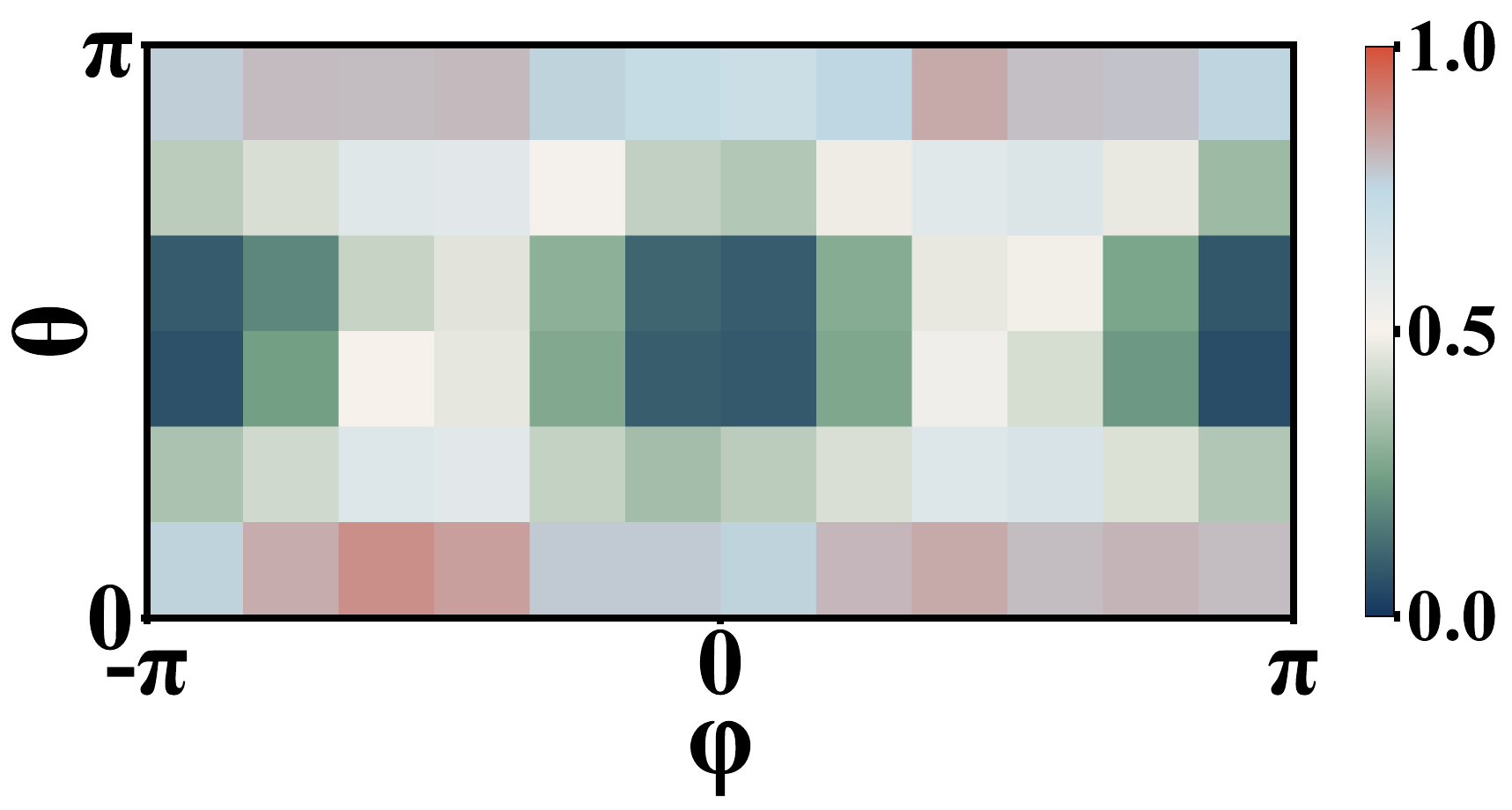}} \\
			\subfloat[$c(1|0, \theta_w, \phi_l)$, epoch 20]{\includegraphics[width=6cm]{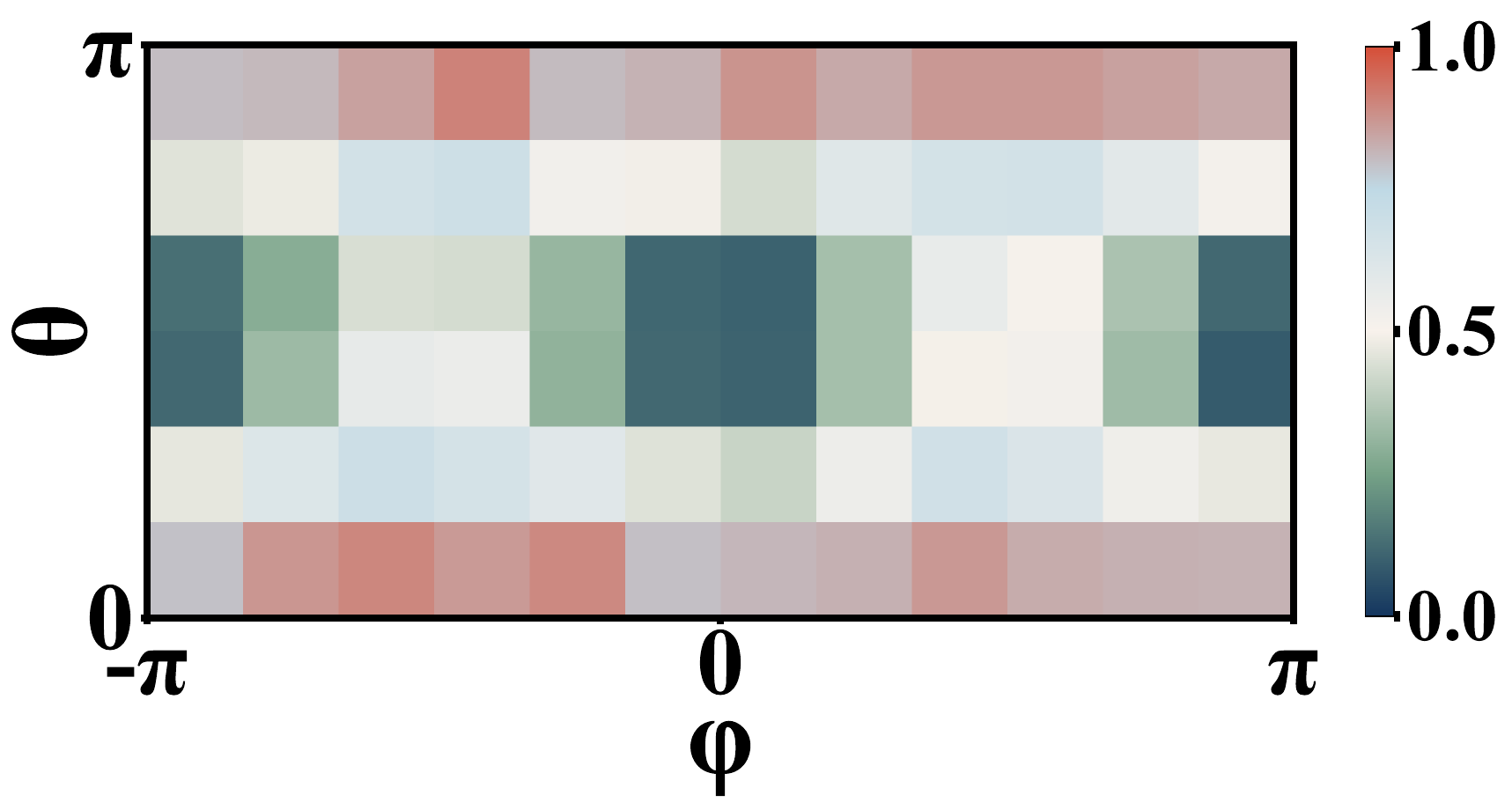}} &
			\subfloat[$c(1|0, \theta_w, \phi_l)$, epoch 100]{\includegraphics[width=6cm]{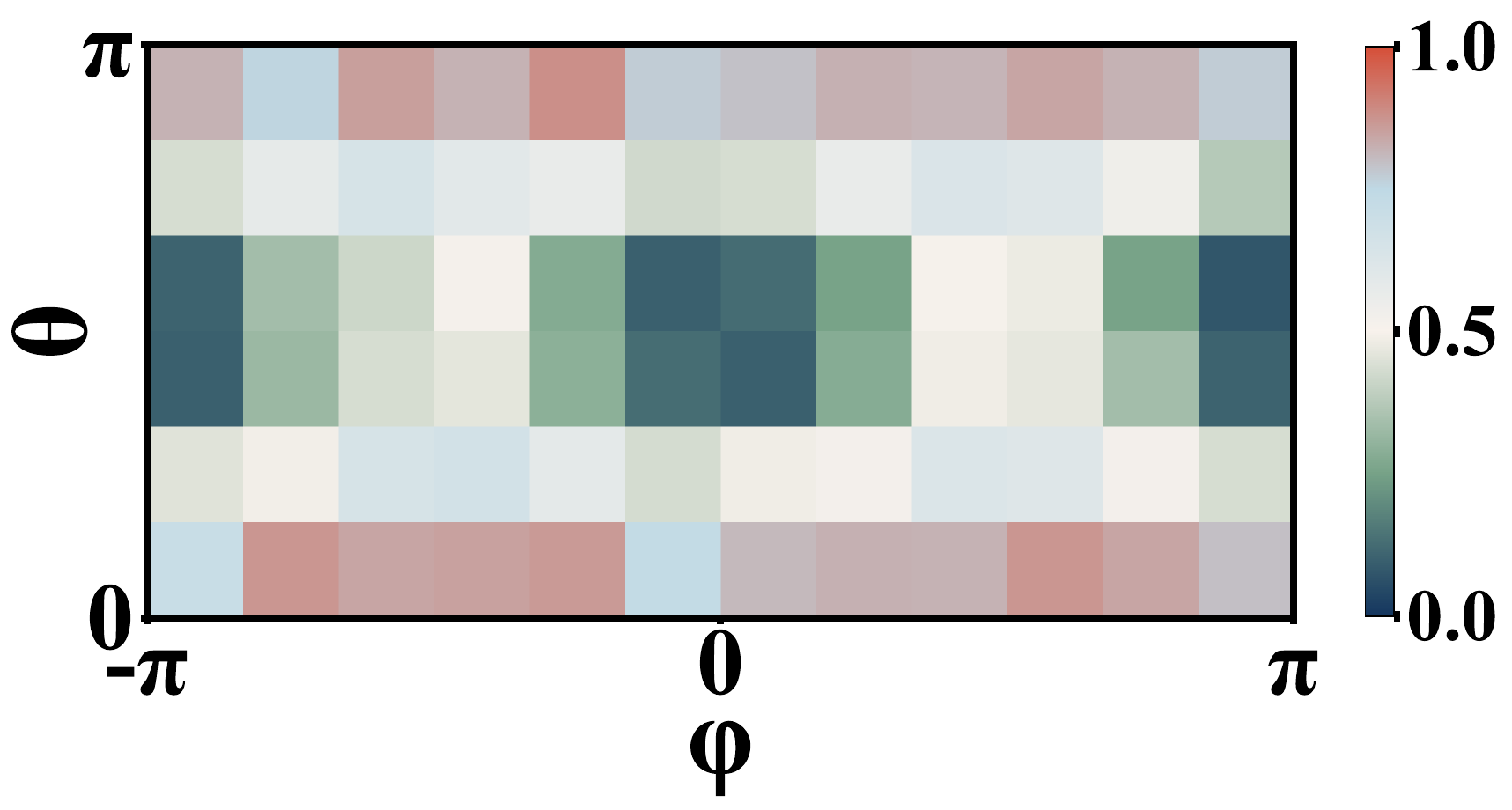}} &
			\subfloat[$c(1|0, \theta_w, \phi_l)$, unfabricated data]{\includegraphics[width=6cm]{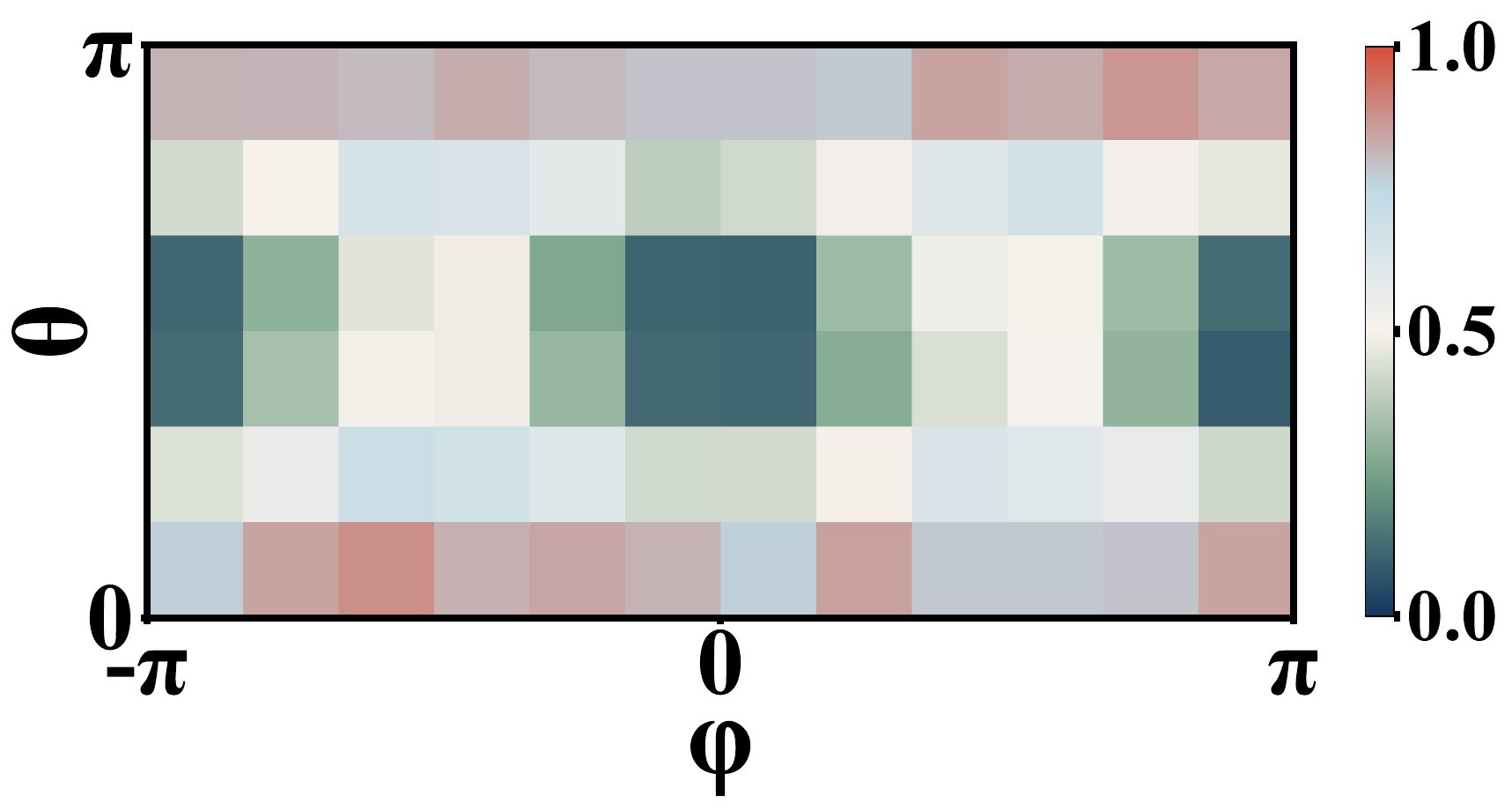}\label{real_experiment01_unfabricated_data}}\\
			\subfloat[$c(0|1, \theta_w, \phi_l)$, epoch 0]{\includegraphics[width=6cm]{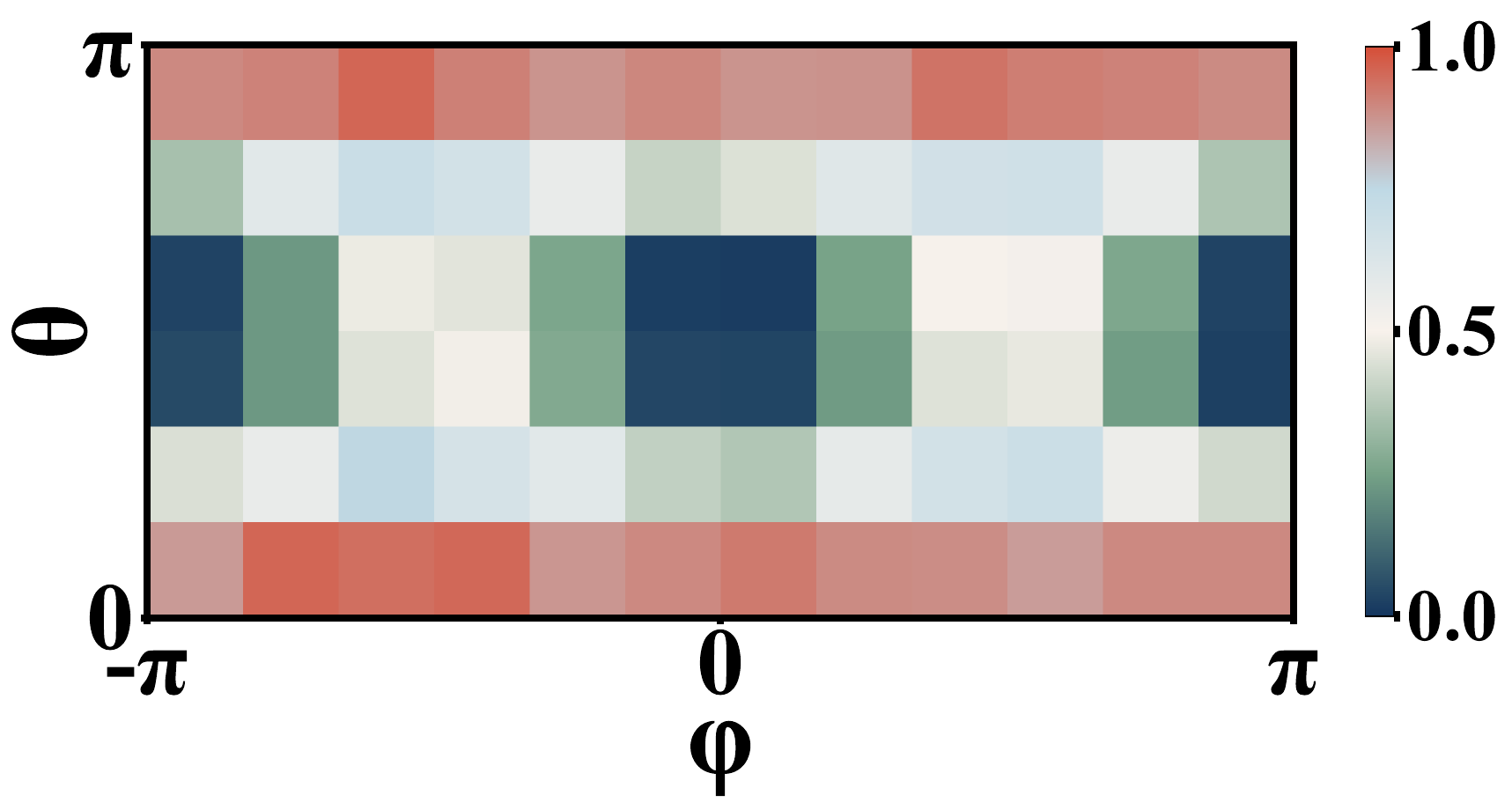}\label{real_experiment10_epoch0}} &
			\subfloat[$c(0|1, \theta_w, \phi_l)$, epoch 1]{\includegraphics[width=6cm]{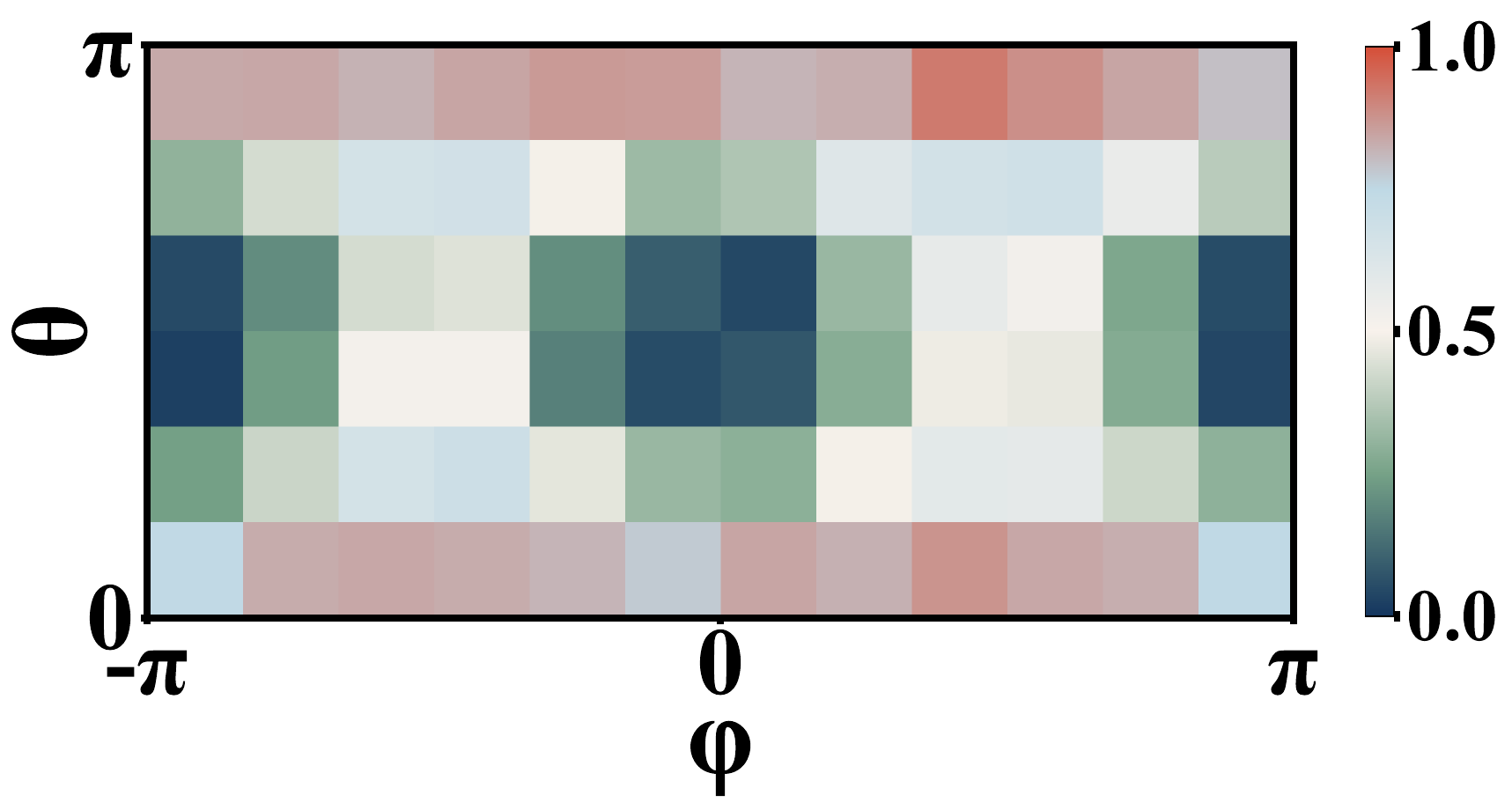}} &
			\subfloat[$c(0|1, \theta_w, \phi_l)$, epoch 5]{\includegraphics[width=6cm]{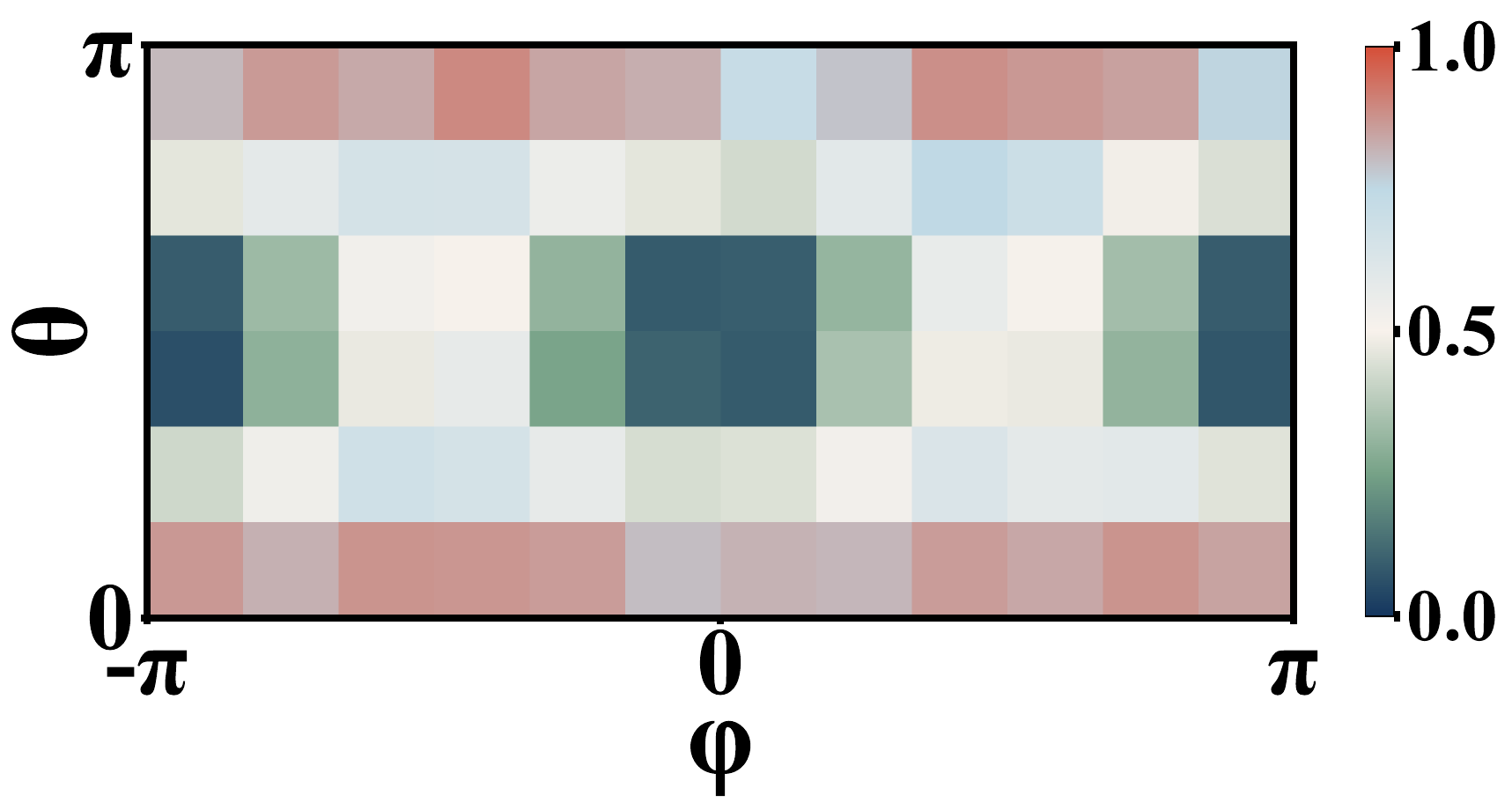}} \\
			\subfloat[$c(0|1, \theta_w, \phi_l)$, epoch 20]{\includegraphics[width=6cm]{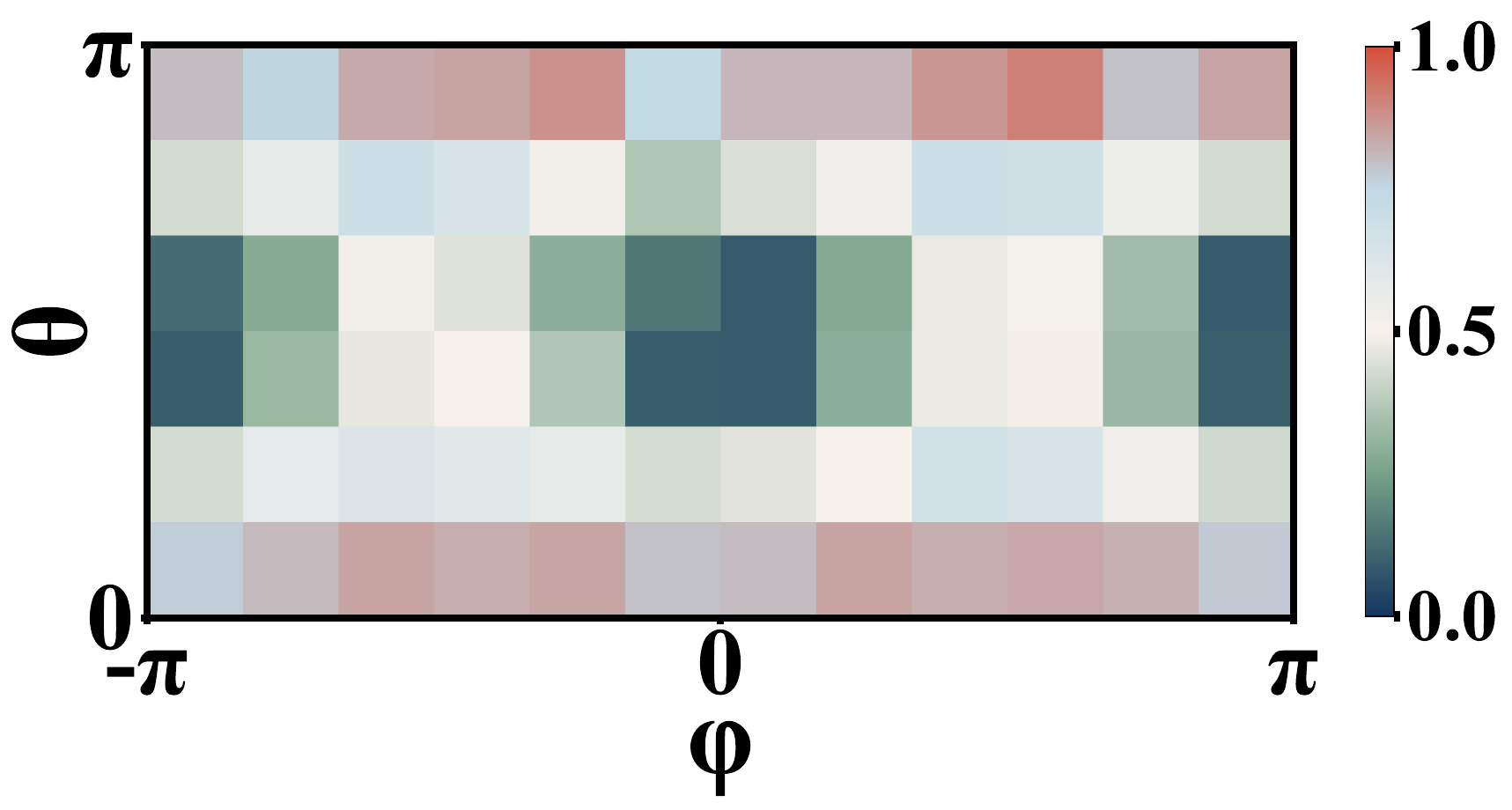}} &
			\subfloat[$c(0|1, \theta_w, \phi_l)$, epoch 100]{\includegraphics[width=6cm]{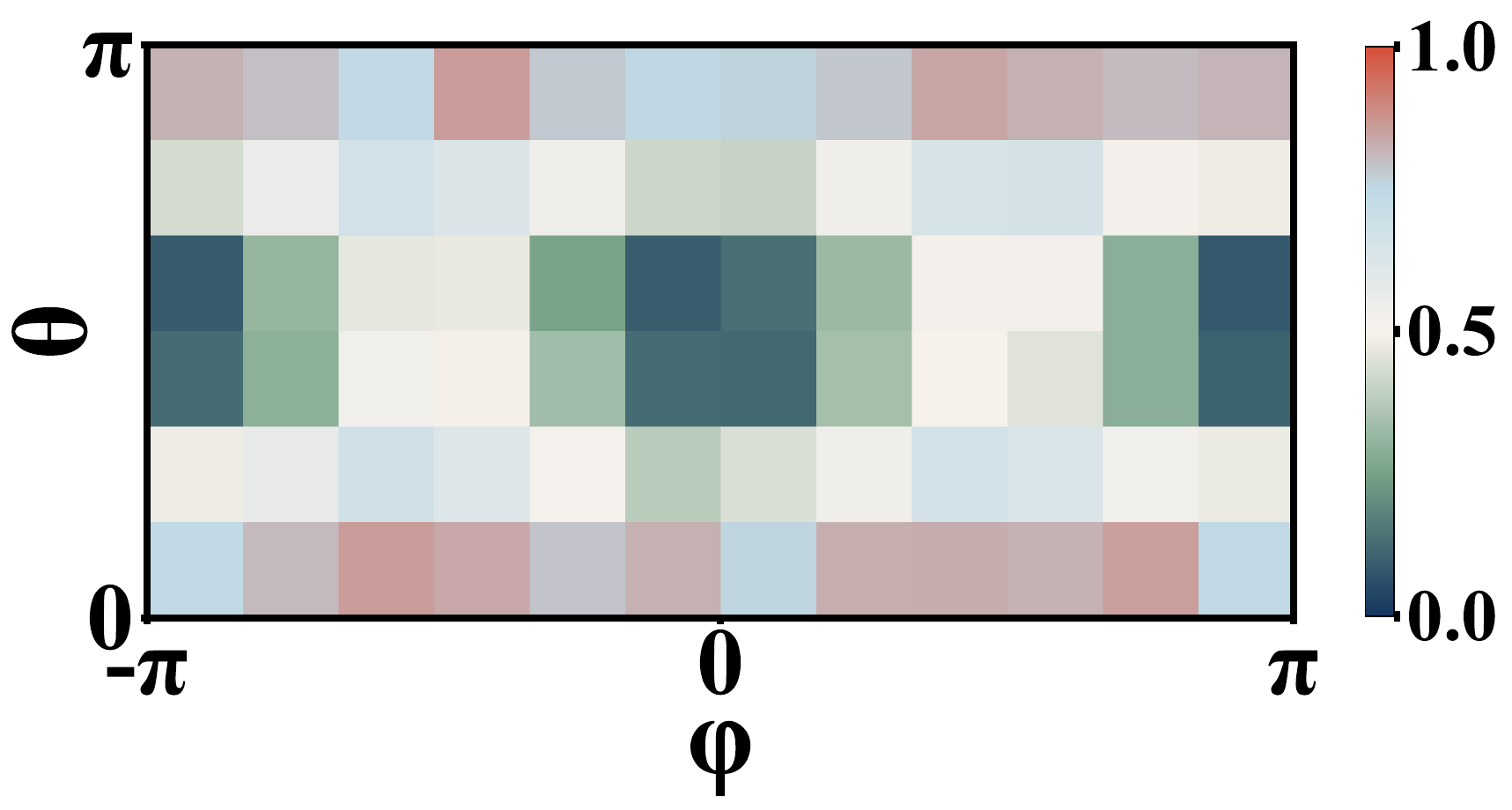}} &
			\subfloat[$c(0|1, \theta_w, \phi_l)$, unfabricated data]{\includegraphics[width=6cm]{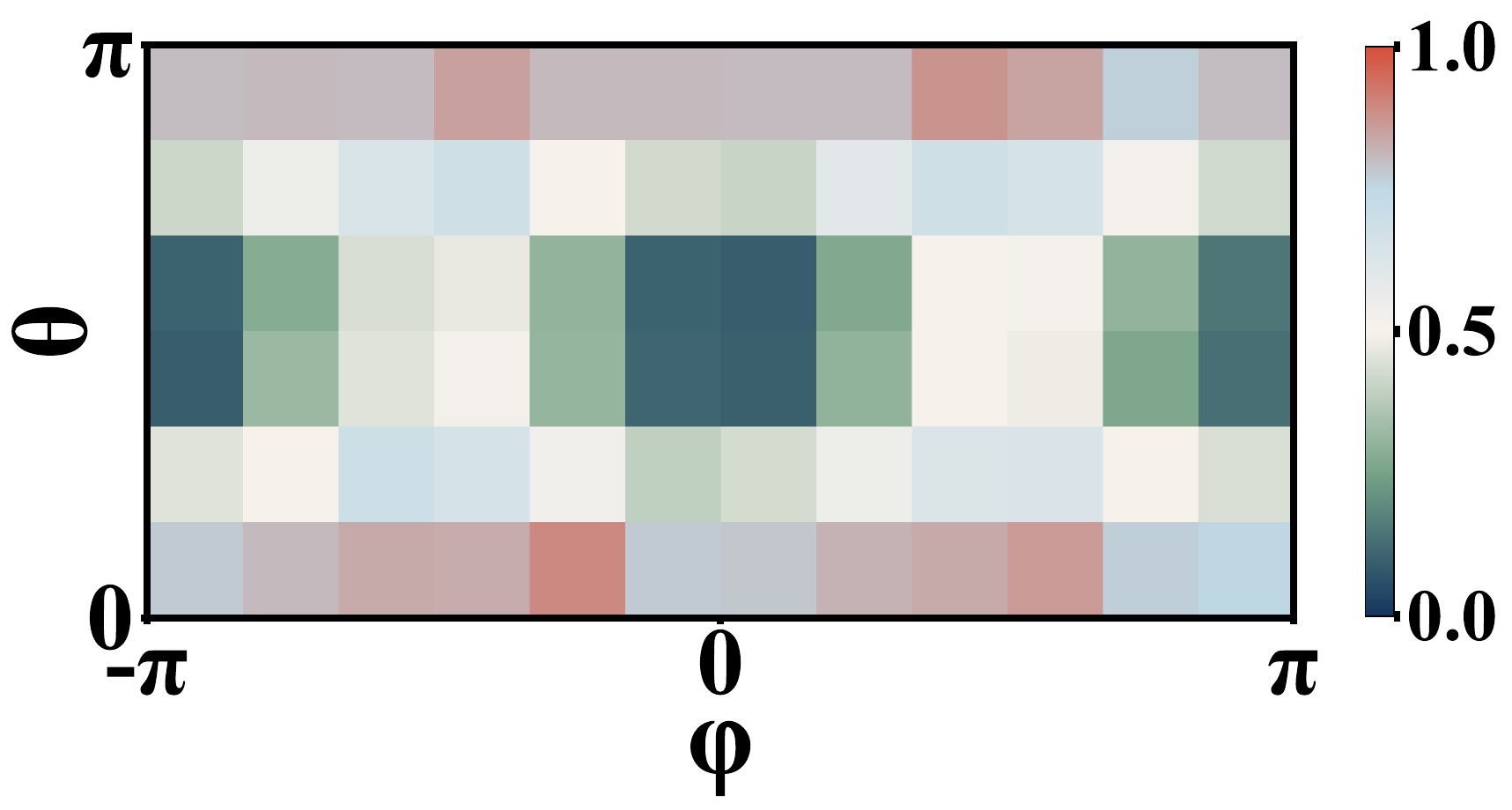}\label{real_experiment10_unfabricated_data}}        \\
			\subfloat[$c(1|1, \theta_w, \phi_l)$, epoch 0]{\includegraphics[width=6cm]{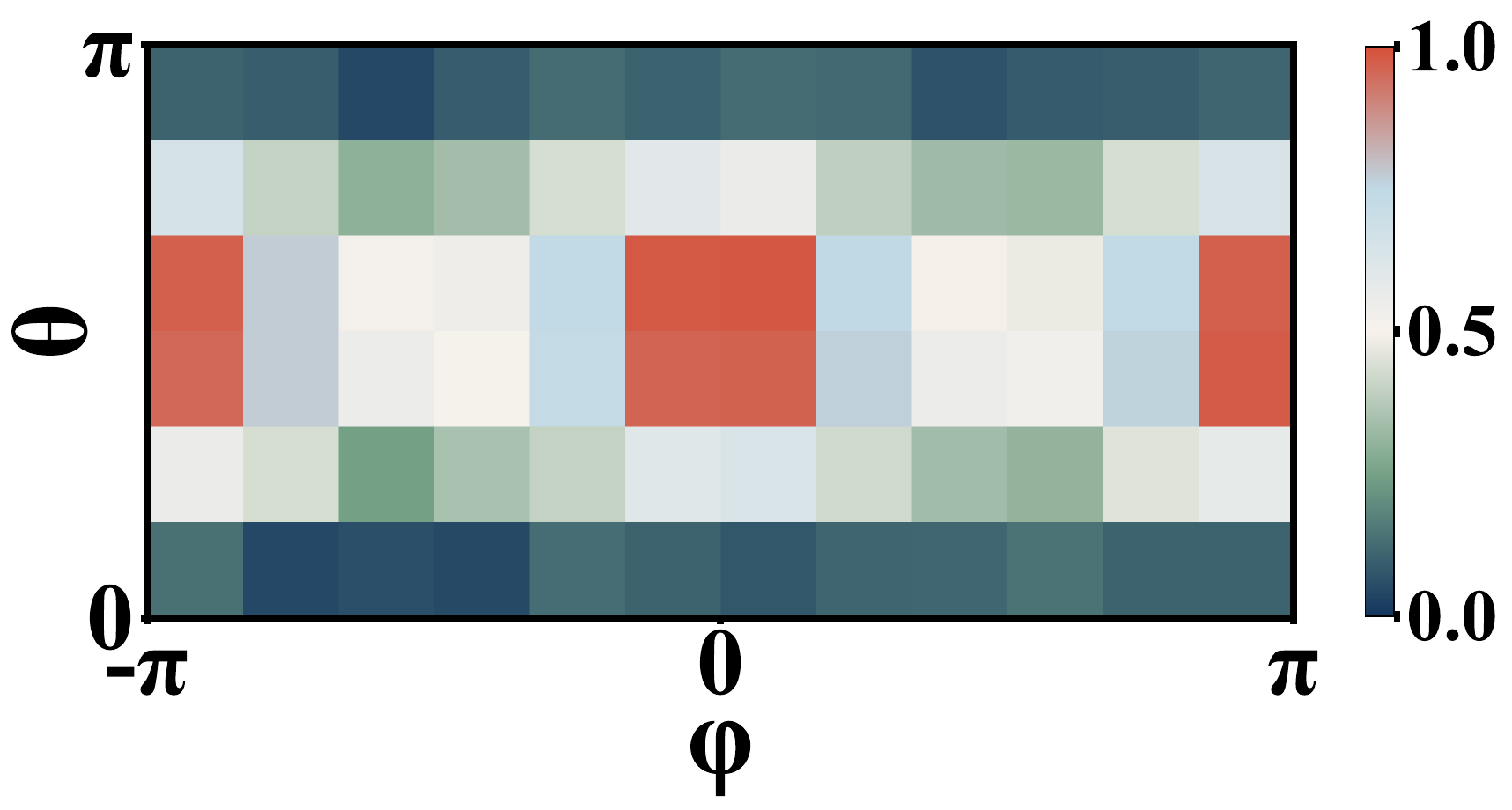}\label{real_experiment11_epoch0}} &
			\subfloat[$c(1|1, \theta_w, \phi_l)$, epoch 1]{\includegraphics[width=6cm]{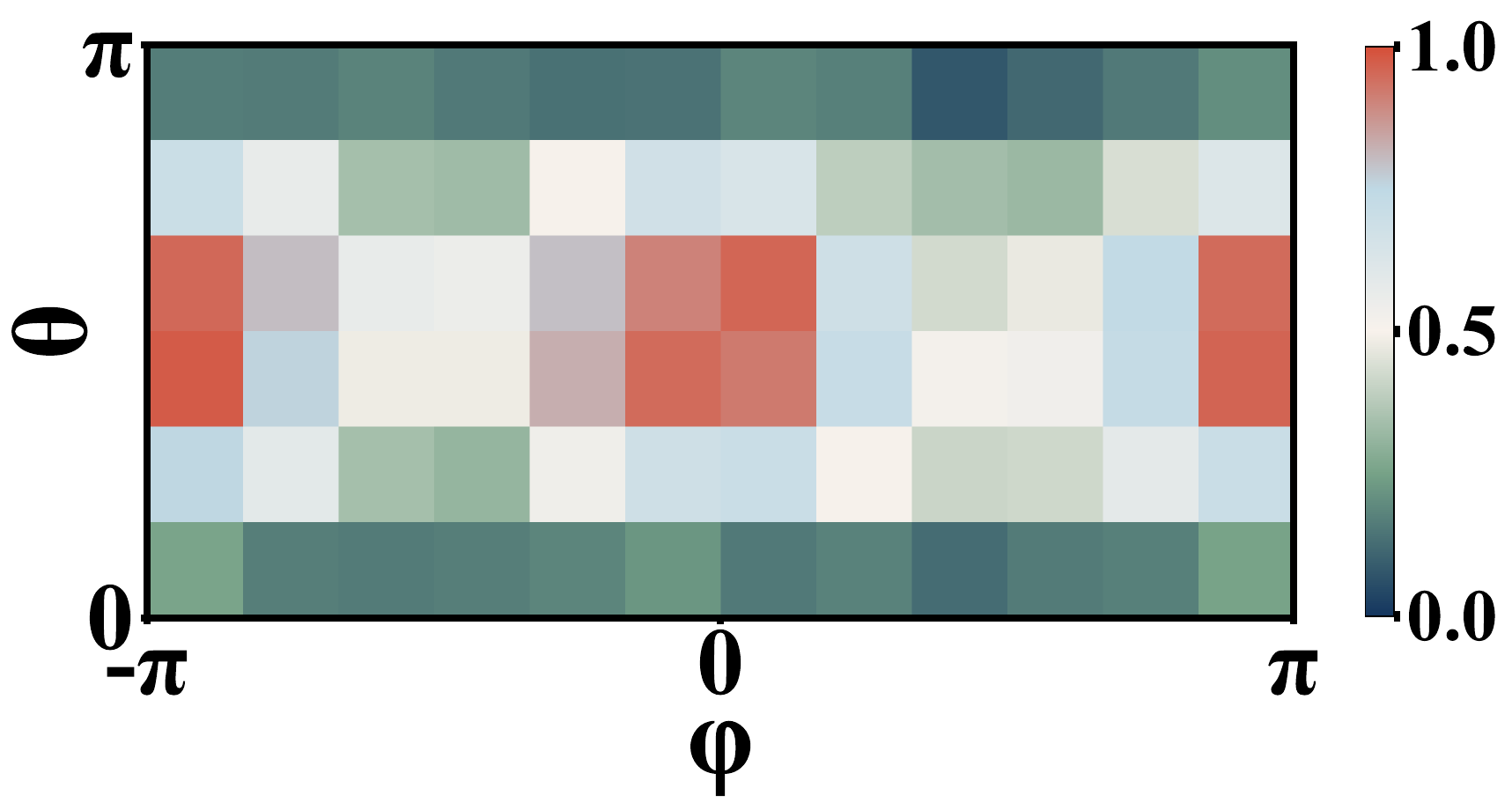}} &
			\subfloat[$c(1|1, \theta_w, \phi_l)$, epoch 5]{\includegraphics[width=6cm]{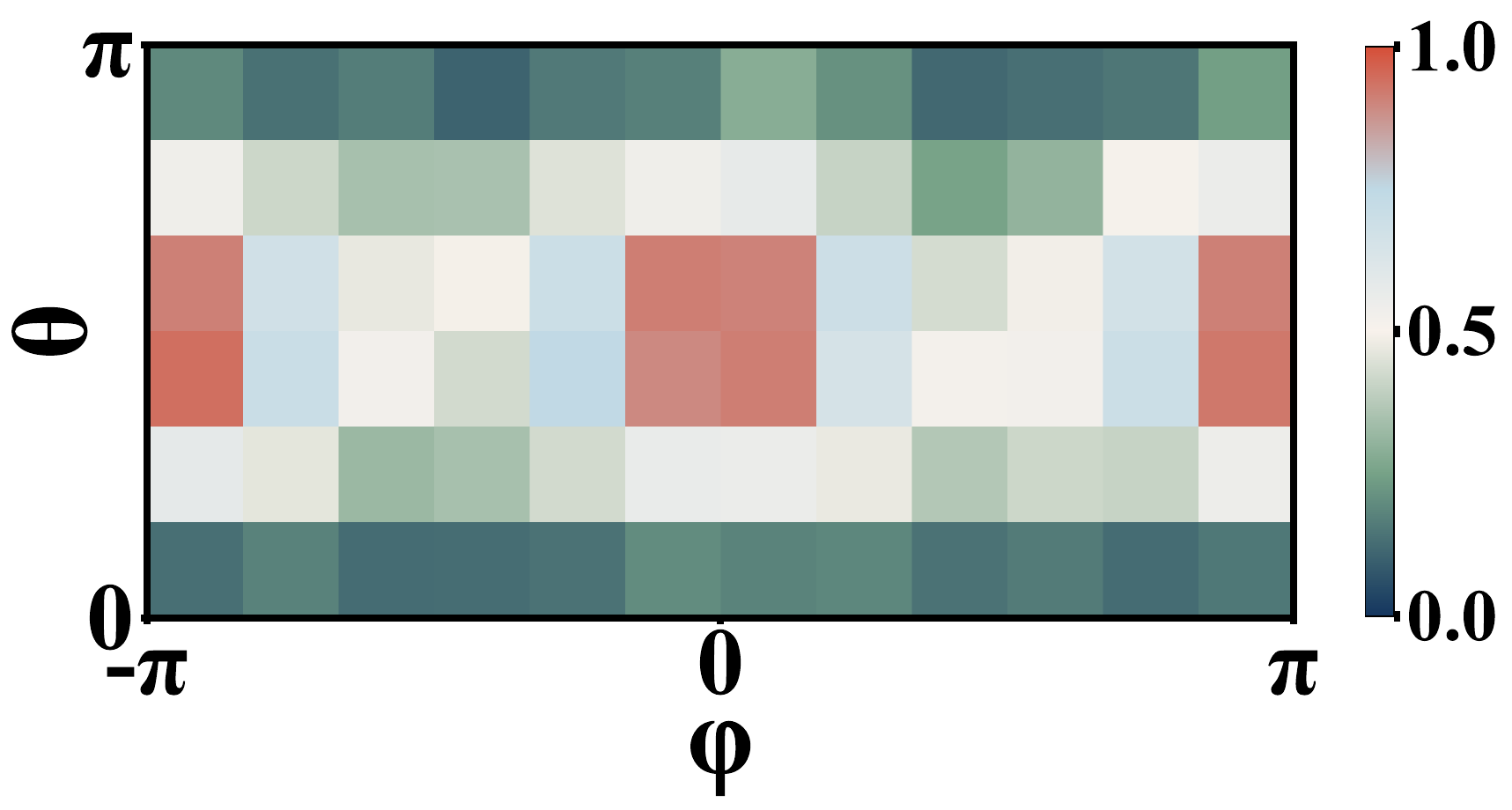}} \\
			\subfloat[$c(1|1, \theta_w, \phi_l)$, epoch 20]{\includegraphics[width=6cm]{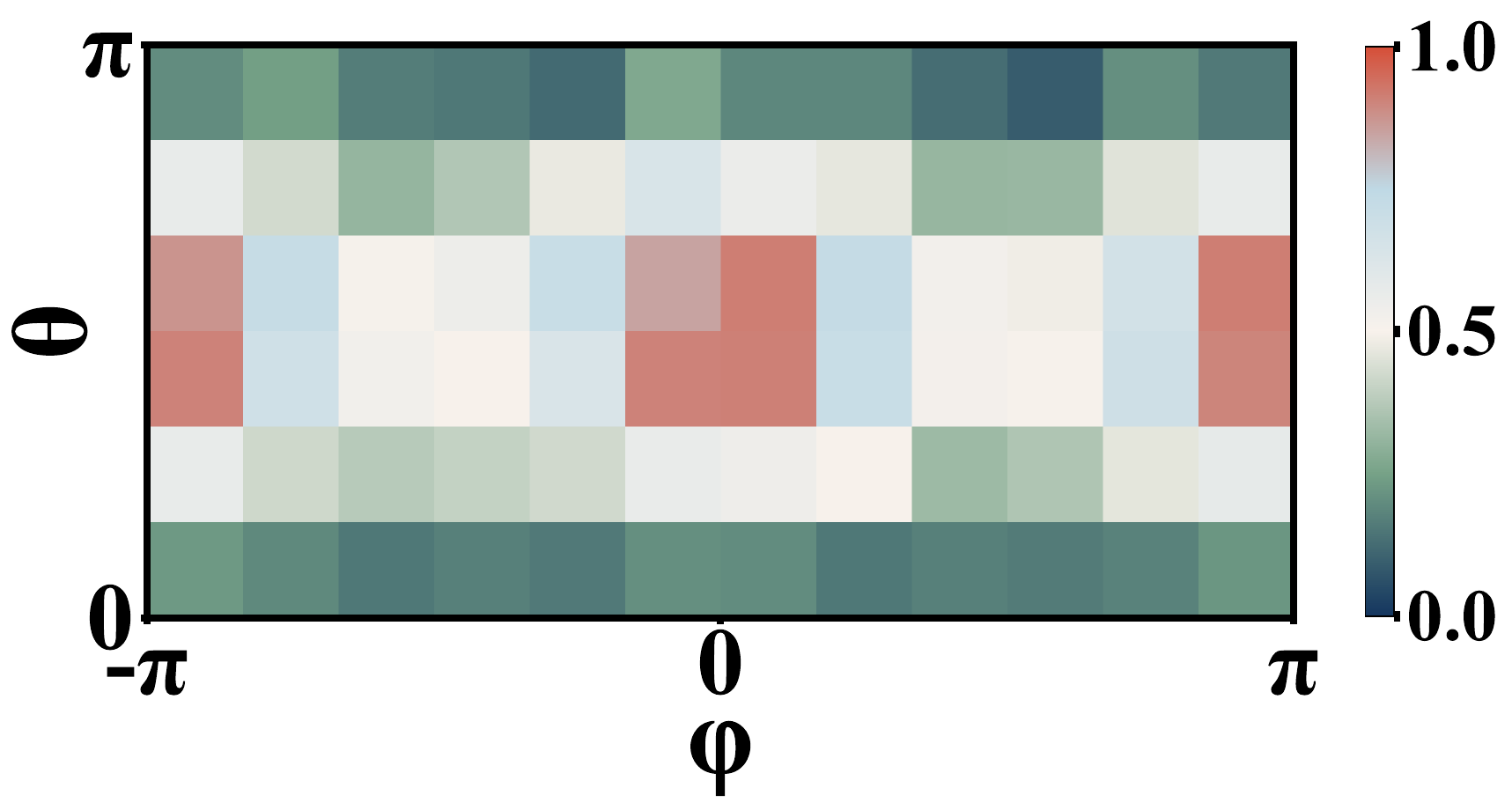}} &
			\subfloat[$c(1|1, \theta_w, \phi_l)$, epoch 100]{\includegraphics[width=6cm]{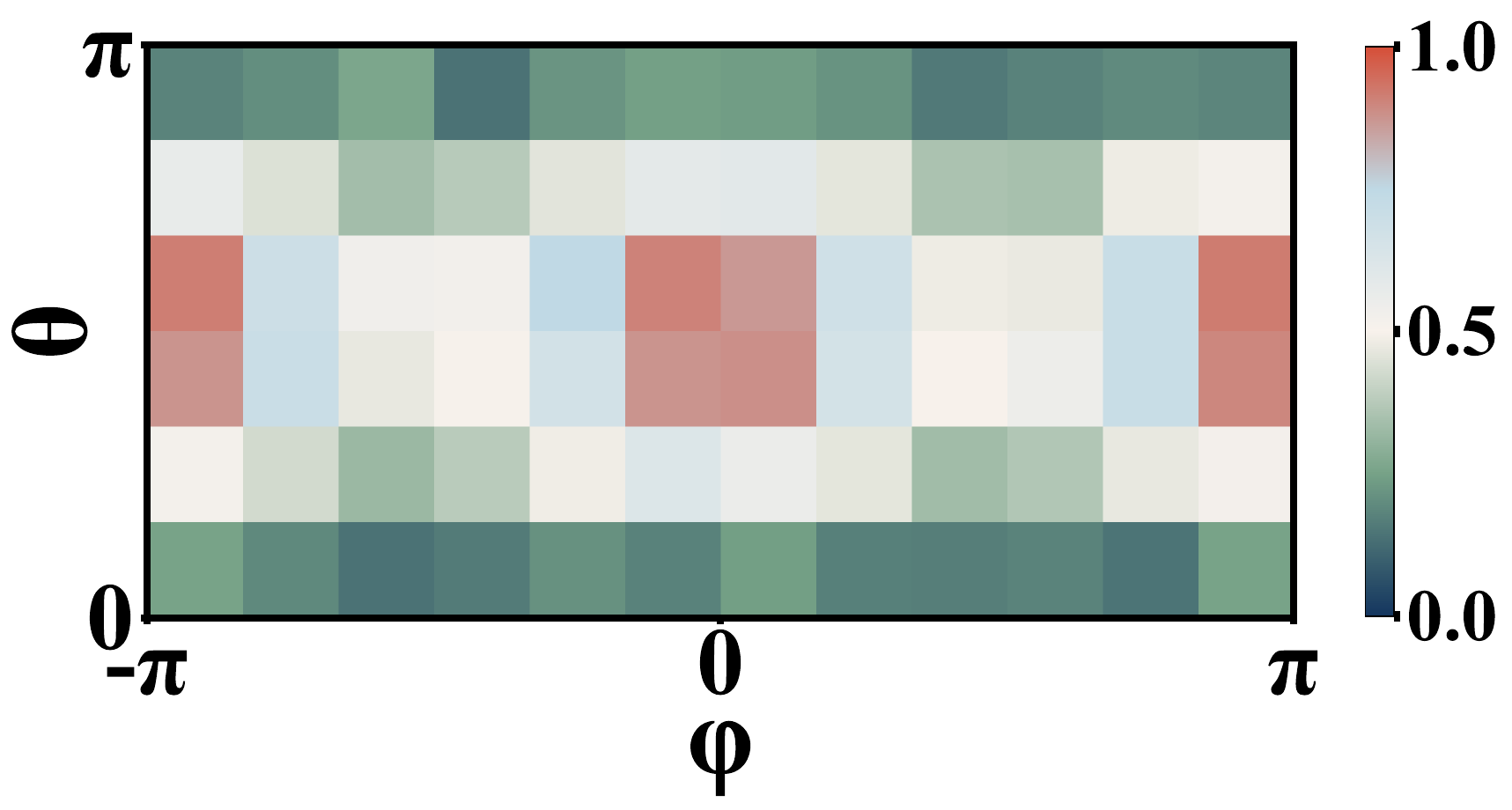}} &
			\subfloat[$c(1|1, \theta_w, \phi_l)$, unfabricated data]{\includegraphics[width=6cm]{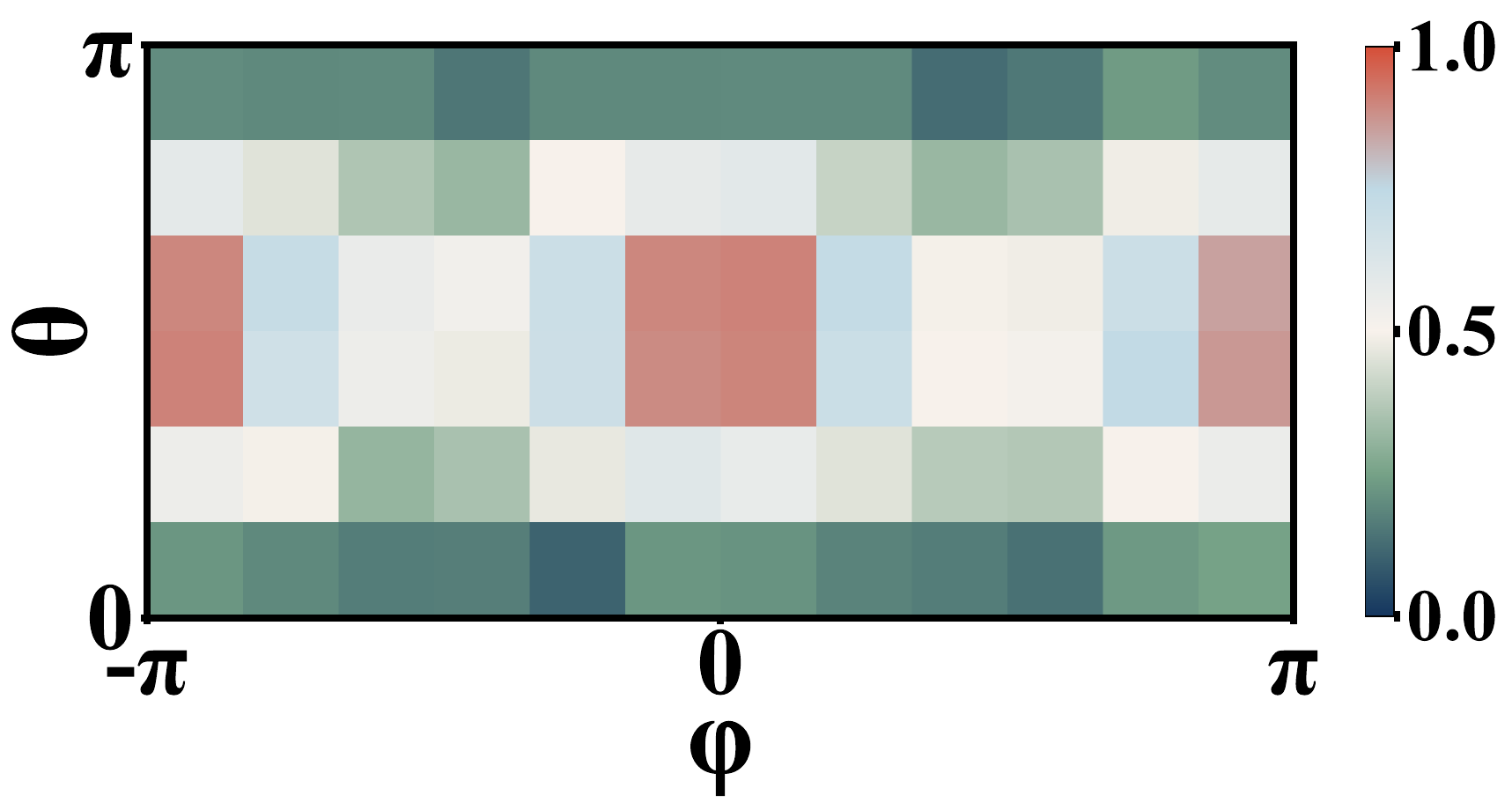}\label{real_experiment11_unfabricated_data}}
		\end{tabular}
		\caption{The distribution of \textit{probability density maps} \(c(b |a, \theta_w, \phi_l)\) is shown for different training epochs and unfabricated data when \(a = 0\) and \(b = 1\) (Figs. \ref{real_experiment01_epoch0}–\ref{real_experiment01_unfabricated_data}), \(a = 1\) and \(b = 0\) (Figs. \ref{real_experiment10_epoch0}–\ref{real_experiment10_unfabricated_data}), and \(a = 1\) and \(b = 1\) (Figs. \ref{real_experiment11_epoch0}–\ref{real_experiment11_unfabricated_data}). The \textit{probability density maps} for \(a = 0\) and \(b = 0\) is plotted in Fig. \ref{real_experiment00} of the main text. }
		\label{real_experiment01}
	\end{figure*}
	The hyperparameters in QP-CNN play a crucial role in minimizing the MSE loss from epoch to epoch. We use the Adam optimizer with an initial learning rate of \(10^{-4}\). The learning rate is adjusted by the optimizer's scheduler, reducing it to \(0.3\) times its original value when the MSE does not decrease for \(2\) epochs. This adjustment helps to lower the loss and prevent instability in the parameters of each neuron. The minimum learning rate is set to \(10^{-12}\). Additionally, the QP-CNN's parameters can be saved in a Python dictionary. We first pre-train the QP-CNN with generated \textit{"uncorrelated"} data, and then train the model with \textit{"correlated"} data. Eventually, the loss function converges by the 100th epoch in QP-CNN.

	\section{Unfitted point clouds in morphing polygons}
\begin{figure}[htbp]
		\centering
		\includegraphics[scale=0.73]{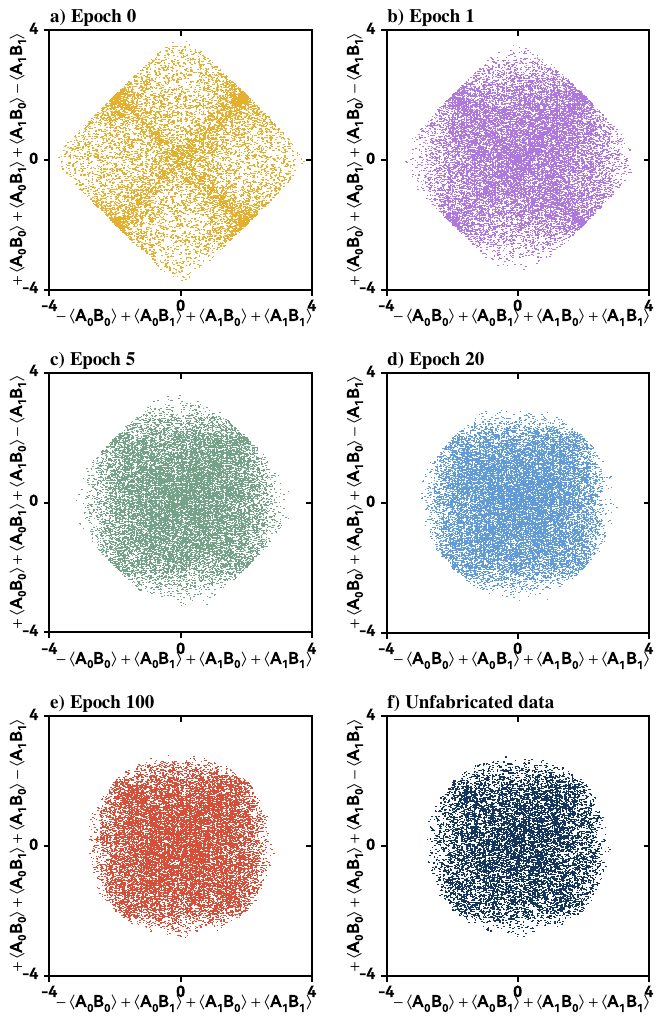}
		\caption{The unfitted point clouds of \textit{morphing polygons} are derived from different epochs and unfabricated data. The coverage of the point clouds represents the varying cognition of Bob, with the main text focusing on the boundaries of \textit{morphing polygons}.}
		\label{point_cloud} 
	\end{figure}

The complete two-dimensional point clouds of \textit{morphing polygons} before the outline points are fitted are shown in Fig. \ref{point_cloud}, with each point computed from Eq. \ref{mp}. The region of the point clouds at different epochs represents the transformation of Bob's cognition. Therefore, we extract the boundary of the point clouds and plot it in Fig. \ref{Bell_polytope}.

	\section{Supplementary data of simulated probability density maps}
    In the simulated \textit{probability density maps}, the result \((a,b)\) can be \((0,0)\), \((0,1)\), \((1,0)\), or \((1,1)\) for each measurement tuple \((x,y)\). In Fig. \ref{real_experiment00}, we show only the case where \(a=0\) and \(b=0\), with the remaining \textit{probability density maps} \(c(b|a, \theta_w, \phi_l)\) shown in Fig. \ref{real_experiment01}. Since \(c(0|a, \theta_w, \phi_l) + c(1|a, \theta_w, \phi_l) = 1\) is obtained by normalizing the exposed events \((b, a, \theta_w, \phi_l)\) with the same \(a\) and different \(b\) in each \((w_{\text{th}}, l_{\text{th}})\) cell, we can observe from Figs. \ref{real_experiment00} and \ref{real_experiment01} that \textit{probability density maps} with identical \(a\) but distinct \(b\) exhibit a complementary relationship.

    %

\end{document}